\documentclass[aps,pra,reprint,superscriptaddress,amsmath,amsfonts,amssymb,longbibliography]{revtex4-1} 
\usepackage{graphicx}
\usepackage{bm}
\usepackage{color}
\usepackage{MnSymbol}
\usepackage{enumerate}
\usepackage{bbold}
\usepackage{lipsum}
\usepackage{array}
\usepackage{textgreek}
\usepackage{amsmath}
\allowdisplaybreaks
\usepackage{mathrsfs}
\usepackage{amsfonts}
\usepackage{graphicx}
\usepackage{mathtools}
\usepackage{multirow}
\usepackage{hyperref}

\usepackage{float}
\usepackage{enumitem}
\usepackage{MnSymbol,wasysym}
\usepackage{fontawesome5}
\usepackage{cancel}
\usepackage{times}
\usepackage{pifont}
\def\be{\begin{equation}}
\def\ee{\end{equation}}
\newcommand{\op}[1]{{\hat #1}}
\newcommand{\dB}{\dd\op{B}}

\newcommand{\psitrue}{\psiful^\text{T}}

\newcommand{\beq}{\begin{equation}}
\newcommand{\eeq}{\end{equation}}
\newcommand{\nn}{\nonumber}

\newcommand{\erf}[1]{Eq.~(\ref{#1})}

\newcommand{\dg}{^\dagger}

\newcommand{\bra}[1]{\langle{#1}|}
\newcommand{\ket}[1]{|{#1}\rangle}

\newcommand{\ito}{It\^o}

\newcommand{\rb}{_{\text{\faFaucet}}}

\newcommand{\ea}{{\em et al.}}

\DeclareMathOperator*{\argmin}{arg\,min}

\definecolor{nblue}{rgb}{0.06,0.3,0.73}
\definecolor{patriarch}{rgb}{0.5, 0.0, 0.5}
\definecolor{nblack}{rgb}{0,0,0}
\definecolor{nred}{rgb}{0.9,0.1,0.1}
\definecolor{nmagenta}{rgb}{0.7,0.0,0.3}
\definecolor{neditcolor}{rgb}{0.3,0.3,0.9}
\definecolor{napricot}{rgb}{0.9,0.6,0.4}

\newcommand{\red}{\color{nred}}

\newcommand{\blk}{\color{nblack}}

\newcommand{\dd}{{\rm d}}
\newcommand{\ddt}{\Delta t}
\newcommand{\dt}{\dd t}
\newcommand{\delt}{\delta t}

\newcommand{\psiful}{\hat{\psi}_{\text{F}; \vec y_t}}

\newcommand{\sechead}[1]{{\em #1}.---}

\usepackage[T1]{fontenc}
\DeclareFontFamily{T1}{calligra}{}
\DeclareFontShape{T1}{calligra}{m}{n}{<->s*[1.44]callig15}{}
\DeclareMathAlphabet\mathcalligra   {T1}{calligra} {m} {n}
\DeclareMathAlphabet\mathzapf       {T1}{pzc} {mb} {it}
\DeclareMathAlphabet\mathchorus     {T1}{qzc} {m} {n}
\DeclareMathAlphabet\mathrsfso      {U}{rsfso}{m}{n}

\makeatletter
\newcommand*\bigcdot{\mathpalette\bigcdot@{.5}}
\newcommand*\bigcdot@[2]{\mathbin{\vcenter{\hbox{\scalebox{#2}{$\m@th#1\bullet$}}}}}
\makeatother

\begin{document}

\title{Quantum trajectories for time-binned data and\\ their closeness to fully conditioned quantum trajectories}

\author{Nattaphong Wonglakhon}
\affiliation{Centre for Quantum Computation and Communication Technology (Australian Research Council), \\ Quantum and Advanced Technologies Research Institute, Griffith University, Yuggera Country, Brisbane, Queensland 4111, Australia}
\author{Areeya Chantasri}
\affiliation{Centre for Quantum Computation and Communication Technology (Australian Research Council), \\ Quantum and Advanced Technologies Research Institute, Griffith University, Yuggera Country, Brisbane, Queensland 4111, Australia}
\affiliation{Optical and Quantum Physics Laboratory, Department of Physics, Faculty of Science, Mahidol University, Bangkok 10400, Thailand}
\author{Howard M. Wiseman}
\affiliation{Centre for Quantum Computation and Communication Technology (Australian Research Council), \\ Quantum and Advanced Technologies Research Institute, Griffith University, Yuggera Country, Brisbane, Queensland 4111, Australia}

\date{\today}

\begin{abstract}
Quantum trajectories are dynamical equations for quantum states conditioned on the results of a time-continuous measurement, such as a continuous-in-time current $\vec y_t$. Recently, there has been renewed interest in finite-interval dynamical maps for quantum trajectories, where the measurement record is available only over intervals of duration $\ddt$. Guilmin \emph{et al.}\ (unpublished) derived such a map for the experimentally relevant case in which only a coarse-grained current (time-binned), denoted by $I_t$, is available. Remarkably, this coarse-grained record $I_t$ still yields a conditioned state, termed the \emph{robinet} state, whose impurity scales as small as $(\Delta t)^{3}$. We show, however, that the \emph{typical} trace distance between the robinet state and the fully conditioned state (obtained from the complete measurement record) scales as $(\Delta t)^{3/2}$. We then introduce a higher-order finite-interval dynamical map, the ``nearly-exact'' map, which requires only one additional real statistic, $\phi_t$, from each interval. We analytically show that the resulting conditioned state is much closer to the fully conditioned state, with a distance scaling as $(\ddt)^{2}$. We further numerically verify the distance scalings, measured relative to highly accurate approximations of the fully conditioned state, for the robinet map, our nearly-exact map, the lowest-order approximate It\^o map, and two existing higher-order maps. Our results show that, for generic systems, whenever $\phi_t$ can be extracted experimentally alongside $I_t$, the nearly exact map provides the most accurate state estimate among the finite-interval maps considered.
\end{abstract}

\maketitle
\date{today}

\section{Introduction}
Continuous quantum measurement~\cite{BookCarmichael,Barchielli1990,Wiseman1993,Wiseman1993-2,Plenio1998} has been the object of sustained attention for decades, especially in quantum optics and, more recently, superconducting circuit experiments~\cite{Murch2013,Weber2014,chantasri2016,Shay2016noncom,BarCom2017,Minev2019,IvaIva2020,Leigh2020,SteDas2022,GuiRou2023}. Such continuous measurements give rise to quantum state evolutions that are stochastic by nature, conditioned on measurement records, known as a \emph{quantum trajectory} or \emph{quantum state filtering}~\cite{Bel99,BookWiseman,BookJacobsSto}. In practice, measurement records are acquired with a finite time resolution, $\ddt$, yielding a \emph{coarse-grained} measurement record or \emph{time-binned current} $I_t \propto \int_t^{t+\Delta t} y_s \dd s$. Here, $y_s$ is a continuous-time process, physically realized at the point of detection, which is not practically extractible in experiment because it would require electronics with infinitesimal time resolution. Standard theoretical descriptions, however, are based on stochastic Schrödinger equations or, more generally, stochastic master equations~\cite{Davies1969,BookGardiner}, both of which use an infinitesimal time step and assume the existence of $y_s$. Consequently, applying these theoretical tools on coarse-grained currents $I_t$ can yield quantum trajectories with significant deviations from the true underlying state evolution.

Several approaches have been proposed to propagate quantum trajectories over finite intervals by means of maps~~\cite{Rouchon2015,Guevara2020,ROUCHON2022,WWC2024,guilmin2025}. Our recent work~\cite{WWC2024} provided an extensive analysis of existing dynamical maps, and introduced a new one with superior properties. Nevertheless, it was only in Ref.~\cite{guilmin2025} by Guilmin \ea\ that a map was introduced that can be calculated (numerically) for arbitrary $\Delta t$ and that describes quantum evolution conditioned solely on $I_t$ by minimizing the Mean Trace-Squared Error between the ``{\em robinet}'' state estimate~\footnote{The subscript we use on $\rho\rb$ was chosen because Ref.~\cite{guilmin2025} call their state the {\em robinet} [faucet or tap] state, a pun on the Franglais term $\rho$-{\em binn\'e} [binned $\rho$].} $\rho\rb$ given the current $I_t$, and the quantum state $\psiful$ (which in this paper we take to be pure) conditioned on the full physical record, $\vec{y}_t$ in each bin. This state, $\rho\rb$, is mixed, even for efficient measurements, because the binning process $\vec y_t \mapsto I_t$ throws away information. However, Ref.~\cite{guilmin2025} also gives an analytic expansion for their map up to $(\Delta t)^{5/2}$, and to this order it generates a state that is pure, so we write it $\hat\psi\rb$.  

\begin{figure}
\includegraphics[width=0.85\linewidth]{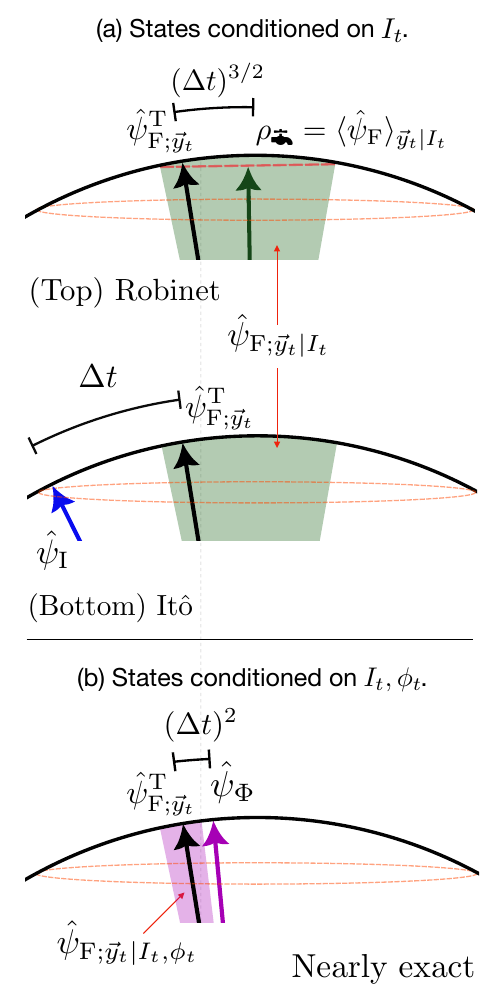}  
  \caption{Bloch representation of pure-state estimates conditioned on different types of records. The hypothetical true state, $\psiful^\text{T}$, is represented by the fixed black arrows. 
  \textbf{(a)} Quantum states conditioned only on $I_t$.
 The possible states conditioned on $I_t$, $\hat\psi_{\text{F}; \vec y_t|I_t}$, are indicated by the light-green shaded area. (Top) The robinet state $\rho\rb$, represented by the green arrow, typically lies at a distance of order $(\ddt)^{3/2}$ from $\psiful^\text{T}$. (Bottom) The crudest estimate, $\hat{\psi}_\text{I}$ (It\^o), is denoted by the blue arrow with a distance of order $\ddt$. \textbf{(b)} Quantum states conditioned on both $I_t$ and $\phi_t$. The possible states conditioned on $(I_t, \phi_t)$, $\hat\psi_{\text{F}; \vec y_t|I_t,\phi_t}$, are indicated by the light-magenta shaded area. The nearly exact state $\hat{\psi}_\Phi$ (magenta arrow) achieves a typical distance of order $(\ddt)^2$ from $\psiful^\text{T}$.}
  \label{fig:Bloch}
\end{figure}

From the fact that the state conditioned on $I_t$ is pure to order $(\Delta t)^{5/2}$ 
one might think that the distance of $\hat\psi\rb$ from the fully conditioned state $\psiful$ would typically be even higher order than $(\Delta t)^{5/2}$. That is not the case, however; the distance is actually typically of order $(\Delta t)^{3/2}$, as depicted in Fig.~\ref{fig:Bloch}(a). This raises the question: can we get a better estimate of $\psiful$, perhaps using an expansion only to order $(\Delta t)^{3/2}$, if we adopt a different binning process? Here we answer that in the affirmative. Instead of binning the current $\vec y_t \mapsto I_t$, its scaled mean in each time interval of duration $\Delta t$, we consider a binning process with two outputs:
\beq 
\vec y_t \mapsto (I_t, \phi_t )  
\eeq
where $\phi_t \propto \int_t^{t+\Delta t} y_s [s-(t+\Delta t/2)]\dd s$. The record $\phi_t$ is a \emph{linearly time-weighted} coarse-grained measurement record. Unlike $I_t$, it captures ``temporal'' structure of the measurement records $\vec y_t$ within each measurement interval in the $\ddt$. \blk  This allows us to introduce what we call a \emph{nearly exact map} ($\Phi$-map), 
an expansion to order $(\Delta t)^{3/2}$ of the map of the fully conditioned state, $\psiful$, which uses both $I_t$ and $\phi_t$. We then show that this $\Phi$-map gives a state, $\hat\psi_\Phi$, that deviates from $\psiful$ only at order $(\Delta t)^2$, i.e., better than $(\ddt)^{3/2}$, as illustrated by Fig.~\ref{fig:Bloch}(b). Moreover, this $\Phi$-map has fewer terms than the robinet method, even when the latter is truncated at order $(\Delta t)^2$. 

We analytically compare the exact (fully conditioned) state, $\psiful$, to the ``nearly exact'' state, $\hat\psi_\Phi$, to the robinet state, $\hat\psi\rb$, and to states evolved by other discrete-time maps that have been proposed in the past, e.g., It\^o~\cite{BookWiseman}, Rouchon-Ralph~\cite{Rouchon2015}, Guevara-Wiseman,~\cite{Guevara2020} and Wonglakhon \ea~\cite{WWC2024}.
We find that quantum states obtained from all existing higher-order maps conditioned solely on $I_t$ typically deviate from the fully conditioned state by an amount of order $(\Delta t)^{3/2}$. However, the size of the deviation depends on the details of the dynamics and the measurement. For example, the robinet method provides an exact map under the conditions of a quantum non-demolition (QND) measurement where the the observed operator $\hat c$ is \emph{normal}, 
so the distance would be limited only by the degree of expansion or other approximation method. We identify several such special cases relevant to the various maps that have been proposed, and verify their performance through numerical simulations. 

By the above methodology, we confirm that the $\Phi$-map, giving a deviation of order $(\Delta t)^2$ from the fully conditioned state $\psiful$, is the best available higher-order map, in general, if one has access to $\phi_t$ as well as $I_t$. This is so whether one is considering an actual (or simulated) experiment with the option of creating a binned record including $\phi_t$ as well as $I_t$ from raw data $\vec y_t$, or whether one is creating a theoretical dual record 
$(I_{t}, \phi_{t})$ for the purpose of simulating an experiment entirely using a finite (but small) time step $\Delta t$.  

The remainder of this paper is organized as follows. In Sec.~\ref{NEM}, we derive the fully conditioned map from the system–bath coupling formalism and review existing approaches. The nearly exact map and the corresponding statistical properties of the measurement records $I_t$ and $\phi_t$ are also introduced. Section~\ref{StateEvo-accuracy} analyzes the error associated with each existing method for quantum trajectory estimation and examines five special measurement cases and their corresponding error magnitudes. Section~\ref{NuExample} presents numerical simulations and analyses of these five measurement scenarios. Finally, we summarize our findings and conclude in Sec.~\ref{conclusion}.

\section{Dynamical maps for quantum trajectory simulations}\label{NEM}

Although standard theoretical tools such as stochastic Schrödinger equations and stochastic master equations are commonly employed for simulating quantum trajectories, these methods generally violate the condition of complete positivity (CP) under a finite time $\ddt$ assumption~\cite{Rouchon2015,ROUCHON2022,WWC2024}. Alternative approaches based on Kraus operators (or \emph{measurement operators}) have been developed, which inherently preserve the CP property ~\cite{BookWiseman,Rouchon2015,Guevara2020,WWC2024}. In this work, we therefore focus on the Kraus operator formalism.

In this section, we first derive the fully conditioned dynamical map from the system–bath interaction formalism in Sec.~\ref{FullConMap}. The fully conditioned state serves as the foundation for constructing various dynamical maps, including the simplest form of It\^o map and some other higher-order maps.  Then, we review the existing higher-order measurement operators in the literature in Sec.~\ref{Comparison_maps}.  Lastly, in Sec.~\ref{NearExMap}, we derive the nearly exact map which is one of the main results in this work.

\subsection{Fully conditioned dynamical map}\label{FullConMap}

To construct the fully conditioned state, let us recall a quantum system undergoing a time-continuous measurement, yielding a continuous result $\vec y_t$ with white noise, as considered in Ref.~\cite{WWC2024}. We will use quantum optics terminology in referring to this as a homodyne measurement and current~\cite{BookWiseman} but note that the same description applies in some non-optical contexts~\cite{Weber2014,Phillippe2015}.  
As is standard, we model this by assuming the quantum system is coupled to a Markovian bosonic field. Given an infinitesimal time resolution $\dt$, the measurement record $y_t$ is collected from time $t$ to $t+\dt$ through homodyne detection. This leads to a Kraus operator defined as 
\begin{align}\label{KrausFull}
    \hat K(y_t)&=\bra{y_t}\hat U_{t+\dt, t}\ket{0},
\end{align} 
where the unitary operator in the rotating-wave approximation takes the form
\begin{align}\label{couplingU}
    \hat U_{t+\dt, t}&=\exp[-i\hat H\dt +\hat c\dB^\dagger_t-\hat c^\dagger\dB_t].
\end{align}
 The Hamiltonian $\hat{H}$ is the system Hamiltonian in the interaction frame, $\hat c$ is a measurement coupling operator, and $\dB_t$ is the bath excitation operator satisfying $[\dB_t, \dB_t^\dagger] = \dt$~\cite{BookWiseman}. The bath is initially in the vacuum state $\ket{0}$, while $\ket{y_s}$ represents the post-measurement state of the bath projected onto an eigenstate of the $\hat Q_t$ quadrature operator, $\hat{Q}_t \ket{y_t} = \dt\, y_t\ket{y_t}$, with $\hat{Q}_t = \dB_t + \dB_t^\dagger$~\cite{WWC2024,BookWiseman}.

To consider a finite-time $\ddt$ evolution, we need to compute a dynamical map for a set of measurement records in the $\ddt$ time bin. Here we define a ``practically infinitesimal'' interval $\delt=\ddt/m$, with $m$ very large, and the list of $m$ results in the interval $\ddt$ by the  vector notation: $\vec y_t=(y_t, \dots, y_{t+(m-1)\delt})^{\rm T}$. The finite-time dynamical map is simply a product of $m$ Kraus operators and taking $m\rightarrow \infty$, leading to the fully conditioned dynamical map given by
\begin{align}
    \hat K(\vec y_t)&=\lim_{m\rightarrow\infty} \hat K(y_{t+(m-1)\delt})\cdots \hat K(y_{t})\nonumber\\
    &=\lim_{m\rightarrow\infty}\prod_{s=t}^{t+(m-1)\delt}\bra{y_s}\hat U_{s+\delt, s}\ket{0},
\end{align}
and the fully conditioned state reads 
\begin{align}\label{fcstate}
    \ket{\psi_{\text{F};\vec y_t}(t+\ddt)}&=\frac{\hat K(\vec y_t)\ket{\psi(t)}}{\sqrt{\wp(\vec y_t|\psi(t))}},
\end{align}
where $\wp(\vec y_t|\psi(t))$ is the probability of getting $\vec y_t$ given that the state before the measurement was $\ket{\psi(t)}$, defined as $\wp(\vec y_t|\psi(t))=\bra{\psi(t)}\hat K^\dagger(\vec y_t)\hat K(\vec y_t)\ket{\psi(t)}$. We refer to the quantum state $\ket{\psi_{\text{F};\vec y_t}}$ as a fully conditioned state because no measurement records are discarded. However, the exponential function in $\hat K(\vec y_t)$ via the definition of $\hat U_{t+\dt, t}$ in \erf{couplingU} is, in general, difficult or impossible to analytically evaluate, and approximations are thus needed for a solvable analysis.

For the analysis below, it is useful to introduce the following notation for the expression of a pure state [e.g., in Eq.~\eqref{fcstate}] as a projector: 
\begin{align}\label{projector}
    \hat\psi_\text{A}(t)&\equiv \ket{\psi_\text{A}(t)} \bra{\psi_\text{A}(t)}.
\end{align}
This allows us to accommodate general scenarios, especially those with a mixed state (in principle), 
such as $\rho\rb$.

\subsection{Existing dynamical maps}\label{Comparison_maps}

We start with approaches existing in the literature. For the sake of simplicity in comparison, we begin by assuming $\hat{H}=0$. Even with this assumption, we can still explore all interesting cases for distinguishing the performance of the different approaches, as we will see in Sec.~\ref{StateEvo-accuracy}. Later in this section we discuss how each existing approach incorporates Hamiltonian evolution.

\sechead{It\^o map} The simplest method is the It\^o measurement operator $\hat M_\text{I}$~\cite{BookWiseman}. It has been shown that the It\^o operator can be derived by approximating $\hat U_{t+\dt,t}$ in Eq.~\eqref{KrausFull} to the lowest order in $\dt$ and using the \emph{It\^o's rule}~\cite{WWC2024}. By projecting to homodyne eigenstates, the resulted map is given by:
\begin{align}\label{ito_mo}
    \hat M_\text{I}(I_t)&=\hat 1+I_t\hat c(\ddt)^{1/2}-\frac{1}{2}\hat c^\dagger\hat c\ddt,
\end{align}
where the dimensionless time-binned record $I_t$ is defined via the coarse-grained records $Y_t$ as
\begin{align}
    I_t&\equiv (\ddt)^{1/2}Y_t,\label{I_t}\\
    Y_t &\equiv \frac{1}{\ddt}\int_t^{t+\ddt}y_s\dd s.\label{coarse-grained-Y}
\end{align}
The dimensionless variable makes the order of $\ddt$  explicit, which is particularly useful when comparing maps order by order, and it is also consistent with Ref.~\cite{guilmin2025}.
Here, their \emph{ostensible probabilities}~\cite{BookWiseman} are given by
\begin{align}
  \wp_\text{ost}(I_t)&=\frac{1}{\sqrt{2\pi}}\exp(-I_t^2/2),\label{postI}\\
    \wp_\text{ost}(Y_t)&=\sqrt{\frac{\ddt}{2\pi}}\exp(-Y_t^2\ddt/2).
\end{align}
These ostensible probabilities are auxiliary probability distributions introduced for the purpose of constructing stochastic measurement updates in a mathematically convenient form. It does not represent the physical distribution of measurement outcomes; rather, it is chosen so that the Kraus operators can be written in a normalized form, while the correct physical statistics are recovered through the state update and normalization. Specifically, one can define the Kraus form via $\hat K_\text{I}(I_t)=\sqrt{\wp_\text{ost}(I_t)}\hat M_\text{I}(I_t)$, which obeys the completeness condition 
\begin{align} \label{Icomplete}
\int \dd I_t \hat K\dg(I_t) \hat K(I_t)  &=\hat{1} 
\end{align} 
with error $O[(\Delta t)^2]$, 
while higher-order maps will obey this condition with error of higher-order in $\Delta t$. Note that here we have used the big-O which we will use (along with little-O notation) throughout the paper. We review the rigorous definitions of this notation in Appendix~\ref{O-Notations}. Note that, in our analysis of the quantum state evolution following Eq.~\eqref{fcstate}, the ostensible probability gets canceled out. Thus, we will henceforth only consider the unnormalized form $\hat M$, and refer it as \emph{the} measurement operator. 

\sechead{Rouchon-Ralph map} The It\^o map can easily cause error in simulation as $\ddt$ is finite. Because of this, a higher-order map [Rouchon and Ralph~\cite{Rouchon2015}], was proposed,  using the Euler-Milstein stochastic simulation method~\cite{Milstein1995,Rouchon2015}. They introduced a stochastic correction term from It\^o map, which gives
\begin{align}\label{RR_map}
    \hat M_\text{R}(I_t)&=\hat M_\text{I}(I_t)+\frac{1}{2}\hat c^2(I_t^2-1)\ddt
\end{align}
for stronger convergent of stochastic terms. Recent work by the current authors [Wonglakhon \ea~\cite{WWC2024}] showed that this can also be derived from the system-bath coupling formalism similar to the It\^o map but without invoking the It\^o rule. However, Ref.~\cite{WWC2024} also showed that the above two methods fail to unraveling Lindblad master equation~\cite{Lind1976,LiLi2018} to order of $(\ddt)^2$.  

\sechead{Higher-order completely positive map} To address the error in matching the Lindblad evolution,  Ref.~\cite{WWC2024} proposed correction terms to the Rouchon-Ralph map, giving
\begin{align}\label{w-mo}
    \hat M_\text{W}(I_t)&=\hat M_\text{R}(I_t)-\frac{I_t}{4}(\ddt)^{3/2}\{\hat c,\hat c^\dagger \hat c\}+\frac{1}{8}(\ddt)^2(\hat c^\dagger\hat c)^2.
\end{align}
This is sufficient to produce ensemble average evolution matching the Lindblad evolution with accuracy to $O(\ddt^2)$.

\sechead{Robinet} Lastly, very recent work by Guilmin \ea~\cite{guilmin2025} has introduced a map to describe quantum evolution conditioned on time-binned currents $I_t$, named \emph{robinet}. They present a dynamical map $\mathcal{K}\rb$, for the density operator $\rho$.   This is designed to produce a state $\rho\rb = \argmin \bar{\sigma}^2_{I_t}(\rho) $, where $\bar{\sigma}^2_{I_t}(\rho)$ is the Mean Trace-Squared Error (MTrSE) with the fully conditioned state $\psiful$, given that only the time-binned current $I_t$ is known: 
\begin{align}\label{MTrSE}
   \bar{\sigma}^2_{I_t}(\rho) &= \langle \text{Tr}[(
   \rho-\psiful)^2]\rangle_{\vec y_t|I_t}.
\end{align}
Here the notation $\langle\bullet\rangle_{\vec y_t|I_t}$ means $\int\dd \vec y_t\wp(\vec y_t|I_t)\bullet$, and $\wp(\vec y_t|I_t)$ is the probability distribution of $\vec y_t$ that gives $I_t$.
From the properties of the MTrSE, the action of $\mathcal{K}\rb$ corresponds to taking the conditional average over all possible measurement records $\vec y_t$ that yield the same $I_t$, i.e.,
\begin{align}\label{robinetexact}
    \rho\rb=\mathcal{K}\rb\hat\psi(t)&=\langle\psiful(t+\ddt)\rangle_{\vec y_t|I_t}.
\end{align}
This results in a mixed state, in principle. However, we can consistently express their map expansion to $O[(\ddt)^2]$ via the measurement operator form as
\begin{align}
    \mathcal{K}\rb &= \wp_\text{ost}(I_t)\hat M\rb(I_t)\bullet \hat M^\dagger\rb(I_t)+O[(\ddt)^{5/2}],
\end{align}
and the measurement operator reads:
\begin{align}\label{rb-mo}
    \hat M\rb(I_t) = & \ \hat M_\text{W}(I_t) +\frac{(\ddt)^{3/2}}{6}(I_t^3-3I_t)\hat c^3 \nn \\
    & +\frac{(\ddt)^2}{12}\bigg[\frac{1}{2}(I_t^4-6I_t+3)\hat c^4 \nn\\
    & \hspace{3em} -(I_t^2-1)(\hat c^\dagger\hat c^3+\hat c\hat c^\dagger\hat c^2+\hat c^2\hat c^\dagger\hat c)\bigg].
\end{align}
That is, to order $(\ddt)^2$, $\mathcal{K}\rb$ is a purity-preserving map. We note that in the expression 
(\ref{rb-mo}) we keep only terms to $O[(\ddt)^2]$, so it differs from the operator as given in Ref.~\cite{guilmin2025}, which includes terms to $O[(\ddt)^{5/2}]$. This is because truncating at $(\ddt)^{2}$ is sufficient to show our results as we will present in Sec.~\ref{StateEvo-accuracy}.  Notice also that all of the terms in \erf{rb-mo} not in $\hat{M}_\text{W}$ are stochastic (they involve the current). From the properties of \erf{postI}, it is not difficult to see that they do not contribute to the Lindblad evolution up to order $O[(\ddt)^2]$, as they 
average to zero. This confirms that $\hat{M}_\text{W}(I_t)$ alone is sufficient to reproduce the Lindblad evolution correctly to $O[(\ddt)^2]$. We summarize the properties of the existing maps in Table~\ref{Summary-all}, highlighting their errors in reproducing individual quantum trajectories and average quantum evolution. The latter is a weaker condition, for which higher-order accuracy is already achieved by $\hat M_\text{W}$ and $\hat M\rb$. The former, however, is a more stringent condition, as the smallest achievable error is $O[(\ddt)^{3/2}]$, whereas our goal is to achieve an even smaller error, which is the problem addressed in the present work. 

Up to now we have considered the existing measurement operators without $\hat H$. 
This requires disucssing the different approaches individually. The It\^o and Rouchon--Ralph approaches include the Hamiltonian by simply adding the term $-i\hat H\ddt$ to Eq.~\eqref{ito_mo} and Eq.~\eqref{RR_map}, respectively. (In one example, Rouchon and Ralph also used the additional higher-order term $-\frac{1}{2}\hat H^2(\ddt)^2$ in their map~\cite{Rouchon2015}.) 
For the $W$-map, we did not include Hamiltonian evolution in Ref.~\cite{WWC2024}. The simplest way to do so is again to add the term $-i\hat H\ddt$ to $\hat M_\text{W}$ in Eq.~\eqref{w-mo}. 
If $[\hat H,\hat M_\text{W}]\neq 0$, this leads to a deviation from the exact map at $O[(\ddt)^{3/2}]$, which is the same size of error as already appear in the $W$-map's trajectories. 
Finally, the robinet state $\rho\rb$ is derived via the Lindblad generator $\mathcal{L}$, which includes $\hat H$. Ref.~\cite{guilmin2025} also provides the expansion of the corresponding measurement operator, which is equivalent to the following modification of \erf{rb-mo}: 
\begin{align}\label{robinet_MH}
    \hat M\rb&\mapsto \hat M\rb(I_t)-i\hat H\ddt -\tfrac{1}{2}(\ddt)^{3/2}I_t\left\{\hat{c}, i\hat H\right\}+O[(\ddt)^2].
\end{align}
Since the different treatments of the Hamiltonian first affect the maps at $O[(\ddt)^{3/2}]$, and since, in almost all cases (see Sec.~\ref{SpecialCases}) the errors that arise without $\hat H$ are this size or larger (see Table~\ref{maps_table}), 
it is easier to perform our comparison with $\hat H=0$. This simplification still preserves the key features needed to distinguish the performance of the existing approaches and the approach here.

In Ref.~\cite{WWC2024}, we considered another higher-order method, proposed by Guevara and Wiseman~\cite{Guevara2020}. We omit it from our analysis here because, although that approach satisfies the completeness condition up to order $(\ddt)^2$~Eq.~\eqref{postI}, it yields a quantum state evolution whose accuracy is no better than that of the It\^o method~\cite{WWC2024}. 

\begin{table}[t!]
\setlength{\tabcolsep}{5pt} 
 \renewcommand{\arraystretch}{1.5} 
\begin{tabular}{|l|c|c|c|c|}
	\hline 
	Conditions &  $\hat{M}_{\rm I}$ & $\hat{M}_{\rm R}$ & $\hat{M}_{\rm W}$ &  $\hat{M}\rb$  \\ 
     \hline
	Individual trajectory & $O(\ddt)$ & \multicolumn{3}{c|}{$O
(\ddt)^{3/2}$}  \\  
	\hline
 Lindblad solution & \multicolumn{2}{c|}{$O
(\ddt)^2$} & \multicolumn{2}{c|}{$O[
(\ddt)^3]$}\\
\hline
\end{tabular}
\caption{\label{Summary-all}
Summary of the orders of error in $\ddt$ for existing maps, both at the level of individual trajectories and of the average evolution, where the latter is compared against the Lindblad evolution. Note that $\hat M\rb$ is truncated at $O[(\ddt)^2]$; consequently, its error in reproducing the Lindblad evolution first appears at $O(\ddt^2)$.}
\end{table}

As stated in the Introduction, the quantum state conditioned solely on the current $I_t$ typically suffers from error arising from the loss of information during the time-binning process. Consequently, all of the maps introduced above generate quantum trajectories with only limited accuracy. In later sections, we show that \emph{typical} error scales as $O[(\ddt)^{3/2}]$ for all methods. To establish this result, we require a solvable higher-order reference state via the nearly exact map, which we derive below.

\subsection{Nearly exact map ($\mathbf{\Phi}$-map)}\label{NearExMap}

In this section, we derive the nearly exact measurement operator starting from the system–bath Hamiltonian formalism. To this end, we revisit the operator $\hat H$ in our derivation. Let us recall the operator $\hat K(y_s)$ for an infinitesimal time interval $\dt$. With the same treatment as in Ref.~\cite{WWC2024}, but truncating Eq.~\eqref{couplingU} to order of $(\dt)^{3/2}$, Eq.~\eqref{KrausFull} can be written as 
\begin{align}
    \hat K(y_s)&=\bra{y_s}\hat U_{s+\dt,s}\ket{0} \equiv \hat K_{3/2}(y_s) + O[(\dt)^2],
\end{align}
where we add the subscript `$3/2$' to denote the truncation order. By expressing the operator in the form, $\hat K_{3/2}(y_s)= \sqrt{\wp_\text{ost}(y_s)}\hat{M}_{3/2}(y_s)$, we obtain the approximated infinitesimal operator given by
\begin{multline}\label{infinitesimal_NEM}
    \hat{M}_{3/2}(y_s)= \hat 1+y_{s}\hat c\dt-\left(\tfrac{1}{2}\hat c^\dagger\hat c +i\hat H\right)\dt+\frac{1}{2}\hat c^2(y_{s}^2\dt-1)\dt\\+\frac{1}{6}y_s\left[\hat c^3(y_{s}^2\dt-3)-2\hat c^\dagger\hat c^2-\hat c\hat c^\dagger\hat c -3\{i\hat H, \hat c\}\right]\dt^2.
\end{multline}
Here, the ostensible probability for $y_s$ is given by a Gaussian function: $\wp_\text{ost}(y_s)=\sqrt{\dt/(2\pi)}\exp(-y_s^2\dt/2)$. Again, we construct the nearly exact map for the finite-time interval via the product of $m=\ddt/\dt$ operators: 
\begin{align}\label{NEM_prod}
  \hat M_{\Phi} &= \lim_{m\rightarrow\infty} \hat{M}_{3/2}(y_{t+(m-1)\dt})\cdots \hat{M}_{3/2}(y_{t}).
\end{align}
We provide the full derivation for Eq.~\eqref{NEM_prod} in Appendix~\ref{NEM_derivation} and construct the nearly exact map, given by
 \begin{multline}\label{Phi-mo}
     \hat M_{\Phi}(I_t,\phi_t)=\hat{1}+I_t(\ddt)^{1/2}\hat{c} -\frac{\ddt}{2}\left[ \hat{c}^\dagger\hat{c} - \hat{c}^2\left(I_t^2-1\right)+i\hat H \right] \\ -\frac{(\ddt)^{3/2}}{2} \bigg(I_t\left\{\hat{c}, i\hat H+\frac{1}{2}\hat{c}^\dagger\hat{c}\right\}+\frac{\phi_t}{\sqrt{3}}\left[\hat{c}, i\hat H+\frac{1}{2}\hat{c}^\dagger\hat{c}\right]\\ -\frac{I_t^3-3I_t}{3}\hat c^3\bigg)+\frac{(\ddt)^{2}}{2}\left[\left(\frac{1}{2}\hat{c}^\dagger\hat{c}\right)^2-\hat H^2\right],
 \end{multline}
where we use the subscript $\Phi$ to emphasize the dependence on an additional unitless record variable $\phi_t$, defined by
\begin{align}
   \phi_t& \equiv 2\sqrt{3}(\ddt)^{-3/2}Z_t,\label{phi_t}\\
    Z_t&\equiv \int_t^{t+\ddt}\dd s y_s\left[s-\left(t+\frac{\ddt}{2}\right)\right].
\end{align}
While the record $I_t$ describes a time-averaged (equally coarse-grained) data, the record $\phi_t$ involves a linearly varying weighting over the time interval. We find the ostensible distribution for $\phi_t$ to be 
\begin{align}
    \wp_\text{ost}(\phi_t)&=\frac{1}{\sqrt{2\pi}}e^{-\phi_t^2/2},
\end{align}
which is also a Gaussian function. This result can be understood by discretizing the integral for $\phi_t$:
\begin{align}
\phi_t = 
\lim_{m\rightarrow\infty} 
\sum_{j=1}^{m}y_j \dt
\left[j\dt-(t+\ddt/2)\right],
\end{align}
where each record $y_j$ is distributed according to
$\wp_\text{ost}(y_j)=\sqrt{\dt/(2\pi)}
\exp\!\left(-y_j^2\dt/2\right).$ Since, in the ostensible distribution, each $y_j$ has zero mean, it follows immediately that $\mathrm{E}[\phi_t]_{y_t}=0$. Moreover, using the statistical independence of $y_j$ and $y_{j'}$ for $j\neq j'$, one finds $\mathrm{E}[\phi_t^2]_{y_t}=1$. More fundamentally, $\phi_t$ is a linear combination of independent Gaussian variables $y_j$, and is therefore itself Gaussian, yielding the distribution above. Further details are given in Ref.~\cite{WWC2024}. With similar calculations, one finds that the Gaussian variables $I_t$ and $\phi_t$ are statistically independent and identically distributed under the ostensible distribution. For convenience we list the moments for $X_t\in(I_t, \phi_t)$ that will be useful: 
\begin{align}\label{moments_Iphi}
    \text{E}[X_t]_{X_t}&=0, & \text{E}[ X_t^2]_{X_t}&=1, &\text{E}[X_t^4]_{X_t}&=3.
\end{align}


The derivation of the measurement operator above follows the same fashion as in Ref.~\cite{WWC2024}. The key difference is that Ref.~\cite{WWC2024} focuses only on the ensemble-averaged evolution up to order $(\ddt)^2$. As a result, terms proportional to $\hat c^3$ were omitted, since they do not contribute to the average evolution at this order. Terms involving $Z_t$ in \erf{Phi-mo} were also neglected because $Z_t$ has zero mean and contributes only at order $(\ddt)^3$.

There are two important points to note from the measurement operator in Eq.~\eqref{Phi-mo}. First, the presence of $\phi_t$ relates to non--QND measurement effects, arising from a non-zero commutator of the system's Hamiltonian and the measurement coupling operator, i.e., $[\hat H,\hat c]\ne 0$, or from the 
operator $\hat{c}$ itself being non-normal i.e., $[\hat c, \hat c^\dagger] \ne 0$. Second, we emphasize that no information is discarded up to $O[(\ddt)^{3/2}]$ in $\hat{M}_{\Phi}(I_t,\phi_t)$, so the map's error consequently appears at $O[(\ddt)^2]$. Moreover, those error terms are stochastic only; we include deterministic terms up to order $(\ddt)^2$, which ensures that the completeness relation which replaces \erf{Icomplete}, 
\begin{align} \label{Iphicomplete}
\int \dd I_t\,\dd \phi_t\, \hat K_\Phi\dg(I_t,\phi_t) \hat K_\Phi(I_t,\phi_t)  &=\hat{1}, 
\end{align} 
is obeyed with error $o[(\Delta t)^2]$ (not $O[(\Delta t)^2]$---see Appendix~\ref{O-Notations} for definitions), and similarly the Lindblad evolution.


For all maps presented, we summarize the differences of each existing method and the $\Phi$-map in Table~\ref{maps_table}, by comparing terms in order in $\ddt$, up to order of $(\ddt)^2$. All methods share common terms only up to order $(\ddt)^{1/2}$. Beyond this, additional correction terms appear. For instance, the term $(I_t^2 - 1)\hat{c}^2 \ddt/2$ serves as the post-\ito\  correction for $\hat{M}_\text{R}$, while $-I_t \{ \hat{c}, \hat{c}^\dagger\hat{c} \}  (\ddt)^{3/2}/4 +(\hat c^\dagger\hat c)^2(\ddt)^2/8$ is the the post-Rouchon-Ralph\  correction for $\hat{M}_\text{W}$, and so on. 
For consistency, we set $\hat H=0$ in the $\Phi$-map to match the other cases where we made that simplification.
 \begin{table*}
 \setlength{\tabcolsep}{3pt} 
 \renewcommand{\arraystretch}{1.5} 
 \begin{center}
\begin{tabular}{|c|c|c|c|c|c|c|}
	\hline
	\textbf{Orders} & All maps & $\hat{M}_{\Phi \cap \text{\faFaucet} \cap \rm W \cap \rm R}$ & $\hat{M}_{\Phi \cap \text{\faFaucet} \cap \rm W}$ & $\hat{M}_{\Phi \cap \text{\faFaucet}}$ & $\hat{M}\rb$ & $\hat{M}_{\Phi}$ \\  
	\hline
$(\ddt)^0$ & $\hat 1$ &&&&& \\ [1ex]
 \hline
$(\ddt)^{1/2}$ & $I_t\hat c$ &&&&& \\ [1ex] 
	\hline
$\ddt$ & $-\frac{1}{2}\hat c^\dagger\hat c$ & $\tfrac{1}{2}(I_t^2-1)\hat c^2$ &&&& \\ [1ex] 
	\hline
$(\ddt)^{3/2}$ & & & $-\frac{1}{4}I_t\{\hat c,\hat c^\dagger \hat c\}$ & $\frac{1}{6}(I_t^3-3I_t)\hat c^3$ && $-\frac{1}{4\sqrt{3}}\phi_t[\hat c,\hat c^\dagger \hat c]$ \\ [1ex] 
	\hline
$(\ddt)^{2}$ & & & $\frac{1}{8}(\hat c^\dagger\hat c)^2$ && \begin{tabular}{@{}c@{}} $-\frac{1}{12}(I_t^2-1)(\hat c^\dagger\hat c^3+\hat c\hat c^\dagger\hat c^2+\hat c^2\hat c^\dagger\hat c)$\\ $+\frac{1}{24}(I_t^4-6I_t^2+3)\hat c^4$ \end{tabular} & \\ [1ex] 
	\hline    
\end{tabular}
\end{center}
\caption{Common and correction terms of maps. The measurement operators are defined in Eqs.~\eqref{ito_mo},~\eqref{RR_map},~\eqref{w-mo},~\eqref{rb-mo} and~\eqref{Phi-mo} (setting $\hat{H}=0$ in the last case, for consistency).  Notation: $\hat{M}_{\rm A \cap B}$ is for terms appearing in both $\hat{M}_{\rm A}$ and $\hat M_{\rm  B}$, and so on.}
\label{maps_table}
\end{table*}

\subsection{Dynamical maps via superoperators} \label{sec:DMVS}

We now extend the $\Phi$-map to a more general scenario involving imperfect measurements $\eta<1$, where $\eta$ is a measurement efficiency. In this case, an additional unmeasured decoherence channel is present as a result of loss of information.  For this, rather than a measurement operator, we need a superoperator, denoted by $\mathcal{K}_\Phi$. We first incorporate the inefficiency by replacing $\hat c$ with $\sqrt{\eta}\hat c$ in the operator $\hat M_\Phi(I_t,\phi_t)$ and then derive the superoperator by  expanding  
$\mathcal{K}_\Phi[\bullet] = \wp_\text{ost}(I_t)\wp_\text{ost}(\phi_t)\hat M_\Phi(I_t,\phi_t)\bullet\hat M_\Phi^\dagger(I_t,\phi_t)$. This leads to
\begin{align}\label{NEM_channel}
    \mathcal{K}_\Phi &\propto
    \mathcal{I}+(\ddt)^{1/2}I_t\mathcal{C}+\ddt\left( \mathcal{L}+\frac{I_t^2-1}{2}\mathcal{C}^2\right)\nonumber\\&+\frac{(\ddt)^{3/2}}{2}\left(I_t\{\!\!\{\mathcal{LC} \}\!\!\}+\frac{I_t^3-3I_t}{3}\mathcal{C}^3-\frac{\phi_t}{\sqrt{3}}[\!\![\mathcal{LC} ]\!\!]\right)\nonumber\\
    &+O[(\ddt)^2],
\end{align}
where $\mathcal{K}_\Phi$ is conditioned on the records $I_t,\phi_t$, and the ostensible probabilities for $I_t$ and $\phi_t$ are the same as before. Here $\mathcal{C}[\bullet]=\sqrt{\eta}\hat c\bullet+\bullet\sqrt{\eta}\hat c^\dagger$ appears in the stochastic measurement terms, with a measurement efficiency $\eta$. The Liouvillian is defined as $\mathcal{L}\bullet=-i[\hat H,\bullet] +\hat c\bullet\hat c^\dagger-\frac{1}{2}\left(\hat c^\dagger\hat c\bullet +\bullet\hat c^\dagger\hat c\right)+\mathcal{L'}\bullet$, where $\mathcal{L}'$ contains any other Lindblad terms arising from additional unmeasured coupling to baths.  
Generalizing Ref.~\cite{guilmin2025}, we have used the notations: $\{\!\!\{\mathcal{LC} \}\!\!\}\equiv\mathcal{LC}+\mathcal{CL}$ and $[\!\![\mathcal{LC} ]\!\!]\equiv\mathcal{LC}-\mathcal{CL}$. We stress that the superoperator $\mathcal{K}_\Phi$ generates quantum trajectory accurately at $O[(\ddt)^{3/2}]$. This allows us to determine the error generated by existing maps. For instance, the robinet method differs from $\Phi$-map, by what we call the ``error superoperator'', defined as
\begin{align}\label{lost-info}
    \mathcal{E}\rb\equiv \mathcal{K}_\Phi-\mathcal{K}\rb&=-\frac{\phi_t}{2\sqrt{3}}(\ddt)^{3/2}[\!\![\mathcal{LC} ]\!\!]+O(\ddt^2).
\end{align} 
From this it is clear that $\mathcal{E}\rb$ indeed captures the non--QND effects. This is why, as stated explicitly in Ref.~\cite{guilmin2025}, the robinet method entails the purity loss with a subtle size of $O[(\ddt)^3]$ despite a unity measurement efficiency ($\eta=1$) and no additional decoherence (${\cal L}'=0$).  We can independently show this as follows. In the notation of \erf{projector}, consider an initial pure state $\hat\psi_0$ and the state $\hat{\psi}_\Phi$ under the $\Phi$-map for an interval $\Delta t$, which is also pure. Then $\hat{\psi}_\Phi= \rho\rb + \mathcal{E}\rb\hat\psi_0+O[(\ddt)^2]$, 
and it follows that
\begin{align}
    \hat{\psi}_\Phi^2 &=\rho\rb^2+ \rho\rb(\mathcal{E}\rb\hat\psi_0) + (\mathcal{E}\rb\hat\psi_0)\rho\rb + (\mathcal{E}\rb\hat\psi_0)^2+O[(\ddt)^4].
\end{align}
By taking the ensemble average over $\phi_t$ via the properties in Eqs.~\eqref{moments_Iphi}, this leads to
\begin{align}\label{purity}
   \text{E}\left[\hat{\psi}_\Phi^2 \right]_{\phi_t}&= \rho\rb^2 + \frac{(\ddt)^3}{12}\left([\!\![\mathcal{LC}]\!\!]\rho_0\right)^2 + O[(\ddt)^{7/2}].
\end{align}
By taking the trace over the system, Eq.~\eqref{purity} is evaluated as $\text{Tr}\left\{ \text{E}\left[\hat{\psi}_\Phi^2 \right]_{\phi_t}\right\}=1$ and leads to
\begin{align}
  \text{Tr}[\rho\rb^2]&=1- \frac{(\ddt)^3}{12}\text{Tr}\left[ \left([\!\![\mathcal{LC}]\!\!]\rho_0\right)^2\right] + O[(\ddt)^{7/2}],
\end{align}
with the subtle effect appearing in order $(\ddt)^3$.

Turning to other existing maps, we derive the maps' superoperators in a similar manner as above. Here, the It\^o approach yields: $\mathcal{K}_\text{I} \propto \mathcal{I}+(\ddt)^{1/2}I_t\mathcal{C}+O(\ddt)$, which, as expected, is only correct to $O[(\ddt)^{1/2}]$. For the Rouchon-Ralph approach, we can define the superoperator $\mathcal{K}_\text{R} \propto \mathcal{I}+(\ddt)^{1/2}I_t\mathcal{C}+\ddt\left[ \mathcal{L}+(I_t^2-1)\mathcal{C}^2/2\right]$,
which directly yields the second line in Eq.~\eqref{NEM_channel}, and leads to the error, given by
\begin{align}\label{RR-error}
    \mathcal{E}_\text{R}&=\mathcal{K}_\Phi-\mathcal{K}_\text{R} = O[(\ddt)^{3/2}],
\end{align} 
making it the simplest approach to produce quantum trajectories accurately to order $\ddt$. We note that the map $\mathcal{K}_\text{R}$ can also be found in Ref.~\cite{Rouchon2015}. 
Since the Rouchon-Ralph method is the simplest map that is correct at $\ddt$, we shall use it in Sec.~\ref{NuExample} for approximately generating fully conditioned states $\psiful$ via an extremely small time increment. Lastly, $\hat M_\text{W}$ results in $\mathcal{K}_\text{W} \propto \mathcal{K}_\text{R}  +I_t \{\!\!\{\mathcal{LC}\}\!\!\}(\ddt)^{3/2}/2+O[(\ddt)^{3/2}]$ with an additional correction term from $\mathcal{K}_\text{R}$. Like both robinet and Ruchon-Ralph, we find that $\mathcal{E}_\text{W} = O[(\ddt)^{3/2}]$.



\section{State evolutions and their error}\label{StateEvo-accuracy}

In general, error of quantum state estimation depend on the amount of information available from the measurement record. That is, a quantum state that is fully conditioned on the entire measurement record $\vec y_t$ yields the most accurate quantum state estimate. However, when some information is lost and only certain types of measurement records are available (e.g., the coarse-grained records like $I_t$ and $\phi_t$),  the resulting estimation inevitably incurs errors. It follows that a quantum state conditioned only on the time-binned record $I_t$ is generally less accurate than one conditioned on more information, such as both $I_t$ and $\phi_t$. 

In this section, we will explore error in quantum state evolution conditioned on records $X_t\in (I_t, \phi_t)$. 
For simplicity we consider the quantum state evolution for $\hat{H}=0$, with only one decoherence channel $\hat c$ (the measured one), and with unit efficiency measurement. Thus we can use the pure-state maps
\begin{align}
    \ket{\psi_\text{A}(t+\ddt)}&=\frac{\hat M_\text{A}(X_t)\ket{\psi(t)}}{\sqrt{\wp_\text{A}}},
\end{align}
for A $\in\{\text{I, R, W, \faFaucet}, \Phi\}$. We conduct the error analysis via its projector form $\hat\psi_\text{A}$. We first evaluate typical error for an arbitrary $\hat c$ of each method in Sec.~\ref{OrdersOfError} and provide the error analysis for special cases in Sec.~\ref{SpecialCases}.

\subsection{Typical error for an arbitrary measurement process}\label{OrdersOfError}
In this section, we consider typical error, by comparing each estimated state with the fully conditioned state $\psiful$, for an arbitrary measurement described by an arbitrary coupling operator $\hat c$.  Particularly in this subsection, we will consider Trace Absolute Error (TrAE) as an error metric, because it is analytically convenient and allows direct comparison with the analytical results of our companion work~\cite{WWC2024}.
Mathematically, for the case of pure states (which is the case, or very nearly so, for all the estimates we consider), the TrAE is defined as $D_{\text{A,B}}=\frac{1}{2}\text{Tr}  \left[|\hat{\psi}_\text{A}-\hat{\psi}_\text{B}|\right] =\sqrt{1-|\bra{\psi_\text{A}}\psi_\text{B}\rangle|^2}$. Here, we use the notation $D$ as it corresponds to a trace distance. We note that the TrSE is defined as $\sigma^2_{\text{A,B}}=\text{Tr}[(\hat{\psi}_\text{A}-\hat{\psi}_\text{B})^2]=2(1-|\bra{\psi_\text{A}}\psi_\text{B}\rangle|^2)$, where the TrAE and TrSE are thus related as $2D^2_\text{A,B}=\sigma^2_\text{A,B}$, and therefore exhibit the same scaling behavior.

A challenge for calculating the error from the true state is that deriving $\psiful$ in the general case is intractable. To overcome this, we instead use the $\Phi$-map in our analysis, as we have it explicitly. This works for the following reasons. 
For any quantum states, the TrAE defined as $D_{\text{A,B}}$ satisfies the \emph{triangle inequality}~\cite{Wilde2017},
\begin{align}\label{errorineq1}
D_{\text{A,B}} &\le D_{\text{A,C}} + D_\text{C,B}.
\end{align}
 From Eq.~\eqref{errorineq1}, one can derive TrAEs relating  $\hat{\psi}_\text{A}$, 
$\psiful$, and $\hat\psi_\Phi$, as $D_{\text{A, F}} \le D_{\text{A}, \Phi} + D_{\Phi, \text{F}}$ and $
    D_{\text{A},\Phi}\le D_{\text{A,F}}+ D_{\text{F}, \Phi}$. Using the fact that $D_{\Phi, \text{F}}= D_{\text{F}, \Phi}$, we obtain
\begin{align}\label{inequalities}
    - D_{\Phi, \text{F}} &\le D_\text{A,F}-D_{\text{A},\Phi}\le D_{\Phi, \text{F}},
\end{align}
which leads to $  |D_\text{A,F}-D_{\text{A},\Phi}|\le   D_{\Phi, \text{F}}$. Given that the $\Phi$-map produces a state whose distance from the fully conditioned state as claimed in Sec.~\ref{NearExMap} is 
\begin{align}\label{DPhF}
D_{\Phi, \text{F}} &= O[(\ddt)^2],
\end{align}
Eq.~\eqref{DPhF} implies that 
\begin{align}\label{F=Phi}
    |D_\text{A,F}-D_{\text{A},\Phi}|\le   O[(\ddt)^2].
\end{align}
Hence, we can indirectly evaluate the error of $\hat\psi_{\rm A}$ relative to $\psiful$ via $D_{\text{A},\Phi}$ as long as these error magnitudes are bigger than $O[(\ddt)^2]$. As we will see below, this is always the case, in general for an arbitrary $\hat c$.

 Here, we expand the TrAE with respect to $\ddt$ using the Maclaurin expansion to evaluate $D_{\text{A,B}}= \text{Tr}[\hat{\psi}_\text{A} - \hat{\psi}_\text{B}]$ 
and find that the typical TrAEs are: $D_{\text{I},\Phi} = O(\ddt)$ and $D_{\text{A},\Phi} = O[(\ddt)^{3/2}]$ for A $\in\{\text{R, W, \faFaucet}\}$. These allow us to establish the typical error of each method, which are given by
\begin{subequations}\label{errorsize_gen} 
\begin{align} 
D_{\text{I, F}} &= O(\ddt),\label{Ito-typical-err} \\ 
\left. 
\begin{array}{l} 
D_{\text{R, F}}\\ D_{\text{W, F}}\\ 
D_{{\text{\faFaucet, F}}} 
\end{array} \right\} &= O[(\ddt)^{3/2}]. \label{gen-error}
\end{align}
\end{subequations}
We note that although $D_{\text{R, F}}$, $D_{\text{W, F}}$, and $D_{\text{\faFaucet, F}}$ have the same magnitude in $O[(\ddt)^{3/2}]$, their prefactor scalings may differ. Moreover, the results in Eqs.~\eqref{errorsize_gen} may change for some specific details of the measurement coupling operator $\hat c$, which we shall see in the next subsection. 

We remind the reader that we use the pure state $\ket{\psi\rb}$ in our analysis as it is solvable, where $\hat{\psi}\rb =\ket{\psi\rb}\bra{\psi\rb}$ via $\hat{M}\rb$ defined in Eq.~\eqref{rb-mo}. Here, $\hat\psi\rb$ is not equal to the mixed state $\rho\rb$ in Eq.~\eqref{robinetexact}, but they can be related by $\rho\rb= \hat{\psi}\rb +O[(\ddt)^{5/2}]$. However, the typical error for $\rho\rb$ still reads $\text{Tr}|\rho\rb-\hat\psi_\text{F}|=O[(\ddt)^{3/2}]$, following Eq.~\eqref{lost-info}.

To provide an intuitive picture of the error of the results in Eqs.~\eqref{errorsize_gen} and the $\Phi$-map, we consider the evolution of a pure state conditioned on the dual data $I_t$ and $\phi_t$, as illustrated in Fig.~\ref{fig:Bloch}. To see error of a quantum state conditioned on $\vec y_t\mapsto (I_t, \phi_t)$, we first define a ``hypothetical true state'', $\psitrue$, conditioned on complete knowledge of measurement records $y_t$, represented by the black arrows. We use this extra superscript (T) to distinguish it from $\psiful$,  which we treat as a possible true state (conditioned on the full record $\vec y_t$) having a probability distribution conditioned on the observer’s available record. That is, $\psitrue$ corresponds to the actual realization of $\vec y_t$ and is therefore fixed and independent of the observer’s information. The state $\psitrue$ thus serves as a reference for pictorially indicating the typical size of the error generated by different trajectory maps.

Fig.~\ref{fig:Bloch}(a) shows the case where only the measurement record $I_t$ is available, $\vec y_t \mapsto I_t$. In this case, the robinet state $\rho\rb$, represented by the green arrow, lies at the midpoint of the light-green shaded region, as it minimizes the MTrSE for quantum states conditioned on $I_t$. Explicitly, it is given by the ensemble average $\rho\rb = \langle\psiful\rangle_{y_t|I_t}$, and is hence a (very slightly) mixed state. The light-green shaded region represents the standard deviation $\sigma_{\vec y_t|I_t}$ of the distribution of $\psiful$ when conditioned only on the record $I_t$. Here, the standard deviation is defined as the root MTrSE from $\rho\rb$, and will scale in the same way as  $D_\text{\faFaucet, F}$. Specifically, using Eq.~\eqref{lost-info} together with Eq.~\eqref{F=Phi}, we have $\sigma_{\vec y_t|I_t}=O(D_\text{\faFaucet, F}) = O[(\ddt)^{3/2}]$. This is both the size of the shaded region, and the 
typical distance of $\rho\rb$ from $\psitrue$, because $\rho\rb$ is an optimal estimate of $\psitrue$. 

By contrast, for non-optimal estimates, the typical deviation is greater than the size of shaded region. At the most extreme,
$\hat{\psi}_\text{I}$ (blue arrow), the estimate from the It\^o map, lies at a typical TrAE of order $\ddt$ from $\psitrue$, as determined by $D_\text{I, F}$ in \erf{Ito-typical-err}.
Other maps such as $\hat{M}_\text{R}$ and $\hat{M}_\text{W}$, exhibit closer deviations of order $(\ddt)^{3/2}$ from $\psitrue$ as in \erf{gen-error} which are the same scaling as that of $\rho\rb$, but larger. 

In Fig.~\ref{fig:Bloch}(b), additional information from $\phi_t$ is available, $\vec y_t \mapsto (I_t, \phi_t)$, and the corresponding possible state, $\hat{\psi}_{\text{F};\vec y_t|I_t,\phi_t}$, shows a smaller uncertainty denoted by the light magenta region.
The standard deviation for this scenario is conjectured to scale as $O[(\ddt)^{5/2}]$, where one would require a ``super $\Phi$-map'' which is accurate to $(\ddt)^2$ to verify this. 
Crucially, even though the nearly exact map does not give an optimal estimate conditioned on $(I_t, \phi_t)$, the state $\hat{\psi}_\Phi$ 
(magenta arrow) lies at a typical distance only of order $(\ddt)^2$ from $\psitrue$. This reflects the fact that the $\Phi$-map is a much more accurate estimate state, which is possible because it requires extracting twice as much information from the full record.



\subsection{Special cases}\label{SpecialCases}
Eqs.~\eqref{errorsize_gen} provide typical error for an arbitrary operator $\hat c$; however, there are special scenarios such that some maps may change their orders of error depending on the properties of $\hat c$. We emphasize that we consider $\hat H=0$ case, so that $[\hat H,\hat c]=0$.  It is important to note that, although our analysis is restricted to $\hat H =0$ for simplicity, this still includes the most generic case, where the nearly exact map outperforms all the others, as we shall see below. In fact, including $\hat H$ in our analysis does not alter any of our conclusions, except for Case 4, where a non-zero $\hat H$ can affect the TrAE of the map $\hat M\rb$, as we shall see below. 

In this section, we again utilize the $\Phi$-map to evaluate the error for different Cases as follows. We remind the reader that the notation $O[(\ddt)^p]$ means a term upper bounded by a constant times $(\ddt)^p$. This is important to bear in mind to see that all of the results are consistent, even when some Cases are a special case of other Cases. (For example, Case 2 is clearly a special case of Case 3.)


\subsubsection*{Case 1: \texorpdfstring{$\hat c^2\propto \hat 1$}{c squared proportional to identity}}

In this case, $\hat M_\text{I}$ is enhanced relative to the general case because the Rouchon-Ralph correction term (proportional to $\hat c^2$) commutes with other terms. As a result, $D_{\text{A}, \Phi} = O[(\ddt)^{3/2}]$, for A $\in \{\text{I, R, W, \faFaucet}\}$. Using Eq.~\eqref{F=Phi}, 
we can conclude the TrAEs for this case as
\begin{align}\label{case1-error}
\left. 
\begin{array}{l} 
D_{\text{I, F}}\\ 
D_{\text{R, F}}\\ 
D_{\text{W, F}}\\ 
D_{{\text{\faFaucet, F}}} 
\end{array} \right\} & = O[(\ddt)^{3/2}].
    \end{align}
A physical example of Case 1 is a qubit Pauli-measurement, e.g., $\hat c \propto \hat\sigma_{z}$, where the operators are defined in Table~\ref{setting}. Notably, Ref.~\cite{WWC2024} has verified this error analysis analytically.

\subsubsection*{Case 2: \texorpdfstring{$\hat c^2= 0$}{c squared is equal to zero}}

This is actually a special case of Case 1, with the additional  consequences that $\hat M_\text{I}=\hat M_\text{R}$ and $\hat M_\text{W}=\hat M_\text{\faFaucet}$. Following Case 1, the TrAEs read:
\begin{align}\label{case2-scaling}
\left. 
\begin{array}{l} 
D_{\text{I, F}}= D_{\text{R, F}}\\ 
D_{\text{W, F}}= D_{{\text{\faFaucet, F}}} 
\end{array} \right\} &= O[(\ddt)^{3/2}].
\end{align}
The qubit fluorescence measurement is an example of Case 2, with $\hat c \propto \frac{1}{2}(\hat\sigma_{x}-i\hat\sigma_{y})= \hat\sigma_-$. Again, Ref.~\cite{WWC2024} has also verified this analytically.

\subsubsection*{Case 3: \texorpdfstring{$\hat c^3= 0$}{c cubed is equal to zero}}

In this case, $\hat M_\text{W}= \hat M_\text{\faFaucet} + O[(\ddt)^{2}]$. We find that $D_{\text{I},\Phi} = O(\ddt)$, while  $D_{\text{A}, \Phi} = O[(\ddt)^{3/2}]$ for A $\in \{\text{R, W, \faFaucet}\}$. Using Eq.~\eqref{F=Phi}, we then have the TrAEs:
\begin{subequations}
    \begin{align}
D_\text{I, F}&= O(\ddt),\\
\left. 
\begin{array}{l} 
D_{\text{R, F}}\\ 
D_{\text{W, F}} \simeq D_{{\text{\faFaucet, F}}} 
\end{array} \right\} & = O[(\ddt)^{3/2}]. \label{ntdse}
    \end{align}
\end{subequations}
Here the $\simeq$ in \erf{ntdse} means the left and the right sides are the same to leading order in $\ddt$, but different at higher-order terms. As an example of Case 3, we can consider a spin-1 system with $\hat c \propto(\hat S_x-i\hat S_y)/2 = \hat S_{-}$. (We provide the operator $\hat S_-$ in Table~\ref{setting}.) This choice does not fall under Case 1 or 4.

\subsubsection*{Case 4: \texorpdfstring{$[\hat c,\hat c^\dagger]=0$}{commutator of c and c dagger is zero}}

That is, Case 4 is where 
the operator $\hat c$ is {\em normal}, i.e., $[\hat c,\hat c^\dagger]=0$. Since $[\hat H,\hat c]=0$ because we are taking $\hat H=0$, this is the case of QND measurements. In this case, the robinet method provides an exact solution where the nontrivial coarse-grained record (e.g., $\phi_t$) makes no contribution, and the exact state evolution is conditioned solely on $I_t$. As $\hat M\rb$ is the expansion truncated at $O[(\ddt)^2]$, we can directly determine
We also find that $D_{\text{I}, \Phi}= O(\ddt)$ and $D_{\text{A}, \Phi} = O[(\ddt)^{3/2}]$ for A $\in \{\text{R, W}\}$. Thus, the TrAEs read:
\begin{subequations}\label{case4-error}
    \begin{align}
D_\text{I, F}&= O(\ddt),\\
\left. 
\begin{array}{l} 
D_{\text{R, F}}\\ 
D_{\text{W, F}}
\end{array} \right\} & = O[(\ddt)^{3/2}],\\
D_\text{\faFaucet, F}&= O[(\ddt)^{5/2}]. \label{onlybecause}
    \end{align}
\end{subequations}

Again, we remind the reader that $\rho\rb$ gives the exact evolution in this case, i.e., $|\rho\rb-\psiful|=0$, and the nonzero error in \erf{onlybecause} appears only because we have defined $\hat M\rb$ to be the robinet map truncated at $O[(\ddt)^2]$. To give an example for this case, we can take $\hat c \propto \hat\sigma_{x,y,z}$ for the qubit systems or $\hat c \propto \hat S_{x,y,z}$ for the spin-1 systems. We choose the latter in Sec.~\ref{NuExample}, as it an example that does not fall under Case 1. We note that, in this case, unlike the other cases, including $\hat H$ can potentially make a difference, but only for the robinet map. Specifically, if $[\hat H,\hat c]\ne 0$, then Case 4 transforms to Case 5 and its error increases to $D_\text{\faFaucet, F}=O[(\ddt)^{3/2}]$. This can be seen by comparing $\hat M\rb$ in Eq.~\eqref{robinet_MH} and $\hat M_\Phi$ in Eq.~\eqref{Phi-mo}.

\subsubsection*{Case 5: no condition} 
This case is defined as the most general scenario where the TrAEs are given by Eqs.~\eqref{errorsize_gen}. For example, we can take a lowering operator for a spin-3/2 system, $\hat L_-$, defined in Table~\ref{setting}, which does not fall into any of the other Cases.

\section{Numerical Simulation}\label{NuExample}

To verify the TrAEs provided in the preceding section, we conduct numerical simulations for five types of measurement couplings, to reflect all five special cases. Again, we consider $\hat H = 0$ and a single monitored coupling operator $\hat c$ with $\eta=1$. We remark that we choose $\eta=1$ (which results in pure states) only for the sake of simplicity. One can easily extend to a mixed state evolution using the superoperator formalism of Sec.~\ref{sec:DMVS},  and we provide this mixed state formalism in Appendix~\ref{MixedStateQTJ}.  The five examples, with numbers corresponding to the special cases, are listed in Table~\ref{setting}. We remark that Example 1 falls under Case 1, as well as Case 4. It is important to note that, while Examples 1 and 2 have been realized experimentally~\cite{Murch2013,Weber2014,chantasri2016,Shay2016noncom,BarCom2017,Minev2019,IvaIva2020,Leigh2020,SteDas2022,GuiRou2023}, Examples 3 and 4 are more complex models designed to illustrate the properties discussed in Sec.~\ref{SpecialCases}. Experimental realization and validation of these examples remain as research questions.  

\begin{table}[t!]
 \setlength{\tabcolsep}{4pt}
 \renewcommand{\arraystretch}{1}
\begin{center}
\begin{tabular}{|c | c | c|}
	\cline{2-3}
    \multicolumn{1}{c|}{} & \multicolumn{2}{c|}{{\bf Setting}}  \\ \hline \hline
	{\bf Example} & System operator $\hat c$ & Initial state\\  
	\hline
	1 & $\sqrt{\gamma/2}\hat\sigma_z, \ \hat\sigma_z=\begin{pmatrix}
        1&0\\0&-1
    \end{pmatrix}$ 
    & \multirow{2}{*}{$\frac{1}{\sqrt{2}}\begin{pmatrix}
        1\\1
    \end{pmatrix}$} \\ [2ex] 
	\cline{1-2}
  2 & $\sqrt{\gamma}\hat\sigma_-, \ \hat\sigma_-=\begin{pmatrix}
        0&0\\1&0
    \end{pmatrix}$ &  \\ [2ex] 
	\hline
     3 & $\sqrt{\gamma}\hat S_-, \ \hat S_-=\begin{pmatrix}
        0&0&0\\1&0&0\\0&1&0
    \end{pmatrix}$ 
    & \multirow{2}{*}{$\frac{1}{2}\begin{pmatrix}
        1\\ \sqrt{2} \\1
    \end{pmatrix}$} \\ [2ex] 
	\cline{1-2}
     4 & $\sqrt{\gamma/2}\hat S_z, \ \hat S_z=\begin{pmatrix}
        1&0&0\\0&0&0\\0&0&-1
    \end{pmatrix}$ &  \\ [2ex] 
	\hline
   5 & $\sqrt{\gamma}\hat L_-, \ \hat L_-=\begin{pmatrix}
        0&0&0&0\\\sqrt{3}&0&0&0\\0&2&0&0\\ 0&0&\sqrt{3}&0
    \end{pmatrix}$ 
    & $\frac{1}{\sqrt{8}}\begin{pmatrix}
        1\\ \sqrt{3} \\ \sqrt{3} \\1
    \end{pmatrix}$ \\ [1ex] 
	\hline
\end{tabular}
\end{center}
\caption{Details of simulation examples: system's coupling operators $\hat c$ and initial states.}
\label{setting}
\end{table}

We will present the simulation procedure for each special case in Sec.~\ref{SimProcedure} and then analyze the numerical results in Sec.~\ref{Num-results}.

\subsection{Simulation procedures}\label{SimProcedure}

To make a fair comparison, we introduce a ``true'' quantum state as a benchmark for evaluating error generated by the existing methods. We aim to generate our true trajectories with errors that are small compared with the error of finite-$\ddt$ trajectories, so that the former errors can be ignored. In this work we take $\gamma\ddt = 10^{-2}$ for the finite $\ddt$ evolution (i.e., this $\ddt$ would be the experimental time scale for binning the record). We can then achieve our aim by using an extremely small time resolution $\delt$, and a quantum trajectory simulation method which gives error $O[(\gamma\delt)^p]$, for $p$ large enough. Specifically, for the true evolution we use $\gamma\delt = 10^{-4}$  and the Rouchon-Ralph method, $\hat M_\text{R}$, for which $p=3/2$. This is the simplest method that ensures the state error is $O[(\gamma\delt)^{3/2}] = O(10^{-6})$ as determined in Eq.~\eqref{RR-error}. This is adequate for comparison with the finite $\ddt$ evolution with errors as small as $O[(\gamma\ddt)^{5/2}] = O(10^{-5})$, which we expect in all cases since we truncate all the maps at $O[(\Delta t)^2]$. 
For the $\delt$ we use it would not have been adequate to use the \ito\ map $\hat M_\text{I}$, because it would give a typical error of $O[(\gamma\delt)]= O(10^{-4})$, and our $\Phi$ map, $\hat{M}_\Phi$, 
generically has errors as small as $O[(\gamma\ddt)^{2}] = O(10^{-4})$.  

Since true trajectories with $\eta=1$ should all be pure states, we can write the true state evolution conditioned on a ``true'' measurement record $y_{t_j}$ at a time $t_j$:
\begin{align}\label{con_evo_pure}
    \ket{\psi_\text{T}(t_j+\delt)}&=\frac{\hat M_\text{R}(y_{t_j})\ket{\psi(t_j)}}{\sqrt{\wp_\text{T}(y_{t_j})}},
\end{align}
using the operator $\hat M_\text{R}(y_{t_j})$ defined in Eq.~\eqref{RR_map}, replacing $\ddt$ with $\delt$ and $I_t$ with $(\delt)^{1/2}y_{t}$).


\sechead{True evolution with $\delt$}At any time step $t$ with a quantum state $\ket{\psi_\text{T}(t)}$, we construct a true measurement record by combining a deterministic and stochastic parts,
\begin{align}\label{records}
    y_{t}\delt &=\mu_{t} \delt +\delta W, 
\end{align}
where the (state-dependent) deterministic component is given by: $\mu_{t} = \bra{\psi_\text{T}(t)}(\hat c+\hat c^\dagger)\ket{\psi_\text{T}(t)}$, which depends on the observed operator $\hat c$. The stochastic component is described by a Wiener increment $\delta W$, generated with a Gaussian distribution with zero mean and $\delt$ variance. Given $y_t$ in \erf{records}, the quantum state is then updated via Eq.~\eqref{con_evo_pure} and then the process is repeated to obtain the true evolution for a fixed total time $T$. 

\begin{figure}
  \includegraphics[width=\linewidth]{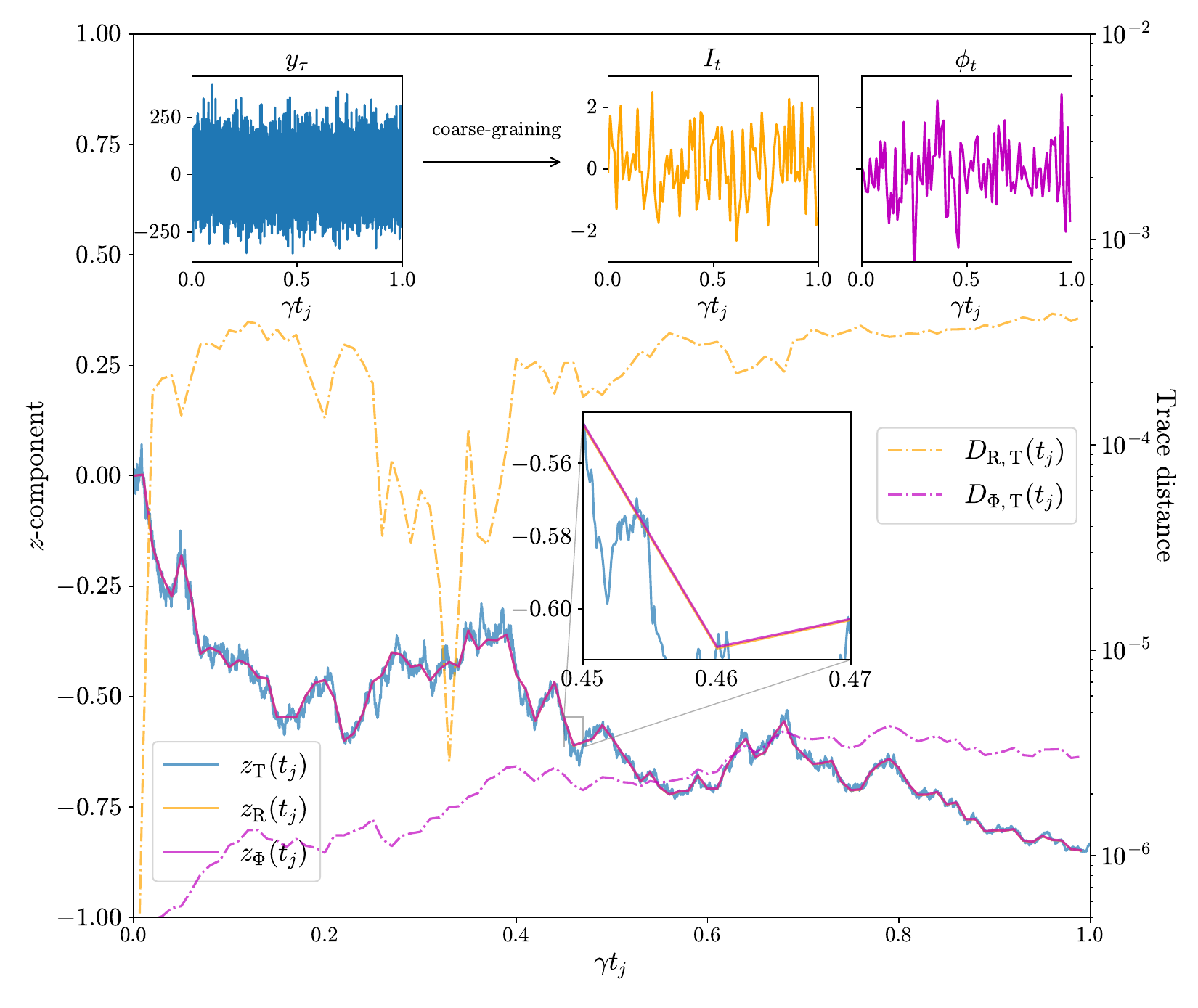}  
  \caption{True and finite-time trajectories. This example depicts the qubit fluorescence 
measurement process with $\hat{c}=\sqrt{\gamma}\,\hat{\sigma}_-$ (Example 2). 
The inset plots show the relevant records: the true record ($y_\tau$) is shown in the first inset in blue, while the coarse-grained records $I_t$ and $\phi_t$ are shown in gold and magenta, respectively. The main plot (left axis) shows the $z$-component trajectories. The true trajectory, $z_{\mathrm{T}}(t_j)=\mathrm{Tr}[\hat{\sigma}_z\hat{\psi}_{\mathrm{T}}(t_j)]$, is shown by the blue curve; the trajectory conditioned on the coarse-grained record $I_t$, denoted by $z_{\mathrm{R}}(t_j)$ and simulated using $\hat{M}_{\mathrm{R}}(I_{t_j})$, is shown by the gold curve; and the trajectory generated by $\hat{M}_\Phi(I_t,\phi_t)$, denoted by $z_\Phi(t_j)$, is shown by the magenta curve. Notice that $z_{\mathrm{R}}(t_j)$ and $z_\Phi(t_j)$ are extremely close. Their differences are therefore illustrated using the trace distances to the true trajectory. The right axis (log scale) shows the corresponding trace distances, $D_{\mathrm{R,T}}(t_j)$ and $D_{\Phi,\mathrm{T}}(t_j)$, indicated by the dash-dotted curves. Parameters: $\eta=1$, $\gamma=1$.}
  \label{fig:Record}
\end{figure}

\sechead{Finite-time evolution with $\ddt$} To simulate the finite $\ddt$ evolution, we start with computing the two coarse-grained records for $y_k\in [t_j, t_j+\ddt]$ by deriving Eqs.~\eqref{I_t} and~\eqref{phi_t}, which are now written in time-discrete forms as
\begin{subequations}\label{cgrecords}
\begin{align}
I_{t_j}\frac{1}{\sqrt{\ddt}} &=\frac{1}{n}\sum_{k=0}^{(n-1)\delt}y_k,\\
\phi_{t_j} \frac{\sqrt{\ddt}}{2\sqrt{3}} &= \frac{1}{n} \sum_{k=0}^{(n-1)\delt}y_{k}(k\delta t -\ddt/2),
\end{align} 
\end{subequations}
using $y_k$ defined in Eq.~\eqref{records}. Here, we used $n=\ddt/\delt=100$ as the number of $\delt$-measurement records in a $\ddt$-time interval $[t_j, t_j+\ddt]$. Once $I_{t_j}$ and $\phi_{t_j}$ are obtained, we then compute the coarse-grained state evolution following 
\begin{align}
    \ket{\psi_{X_{t_j}}(t_j+\ddt)}&=\frac{\hat M_\text{A}(X_{t_j})\ket{\psi(t_j)}}{\sqrt{\wp_\text{A}(X_t)}}
\end{align}
for A $\in \{\text{I, R, W, \faFaucet}, \Phi\}$ and $X_{t_j} = I_{t_j}$ or $(I_{t_j}, \phi_{t_j})$. For the fixed total time $T=10^4\delt=\gamma^{-1}$, we obtain $N=T/\ddt=100$ coarse-grained states per trajectory for each method.  We illustrate a simulated binned trajectory in Fig.~\ref{fig:Record} for the example of a measurement of qubit fluorescence with $\hat c=\sqrt{\gamma}\hat\sigma_-$ (Example 2) using the Rouchon-Ralph map $\hat M_{\rm R}$ as well as the ``true'' trajectory.

\begin{figure*}[t!]
  \centering
  \includegraphics[width=\textwidth]{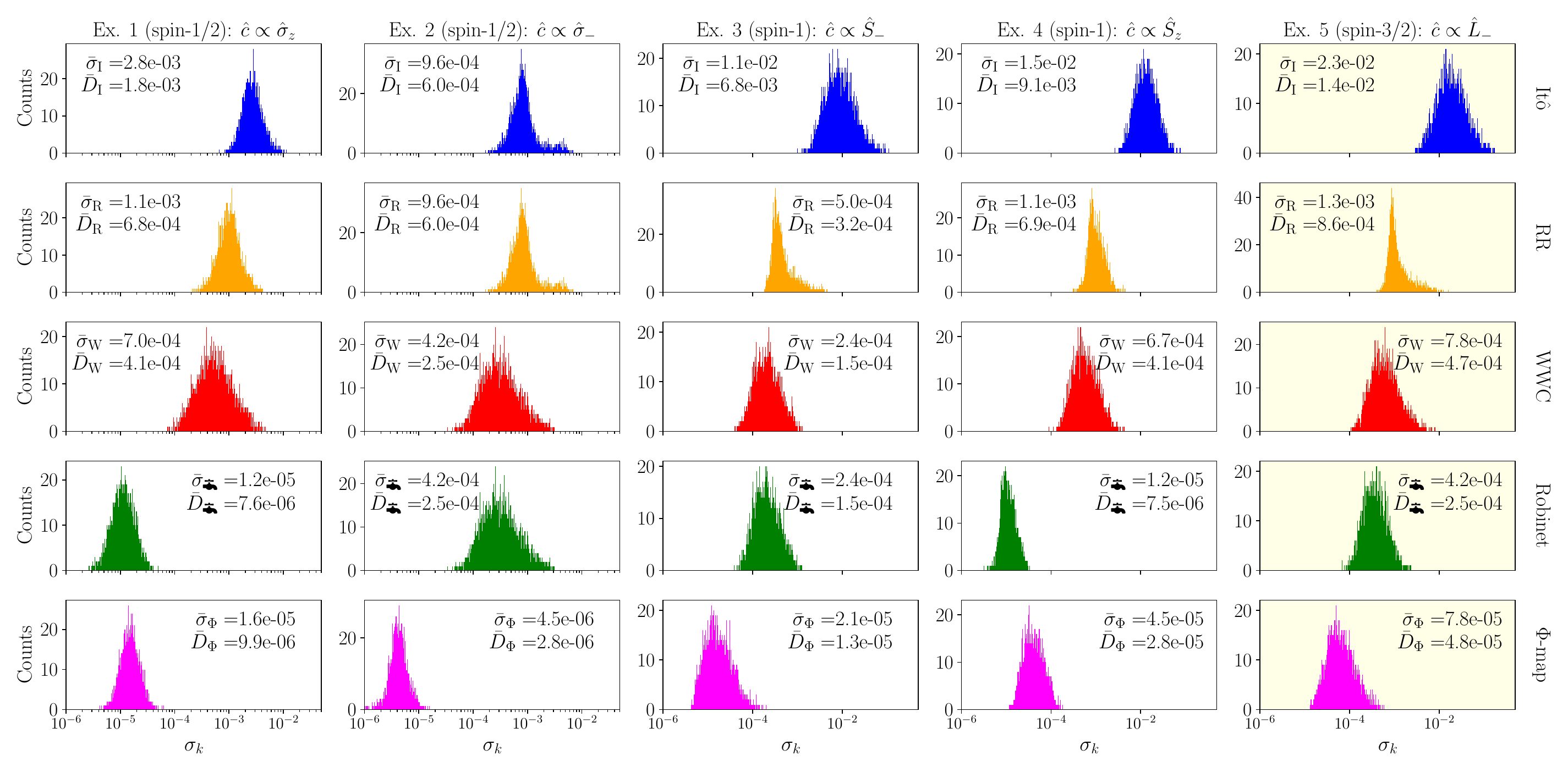}
  \caption{ \textbf{Square root time-average trace-squared errors}: the histograms show $\sigma_{\text{A};k}$ computed via the square root of Eq.~\eqref{TSE1} (log-scaled) of individual trajectories, $\ket{\psi_\text{A}}$, generated by each method comparing with the true state, $\ket{\psi_\text{T}}$, of five special quantum measurement examples. The MTrSE ($\bar \sigma_\text{A}$) and MTrAE ($\bar D_\text{A}$) for method A are reported in the plots. The subscript texts are A $\in\{\text{I, R, W, \faFaucet}, \Phi\}$, used to indicate the numerical results from the maps: $\hat M_\text{I}$ (blue--first row), $\hat M_\text{R}$ (gold--second row), $\hat M_\text{W}$ (red--third row), $\hat M_\text{\faFaucet}$ (green--fourth row) and $\hat M_\Phi$ (magenta--last row), respectively. Parameters: $\hat H=0,$ and $\gamma=1$ for all cases. Note that we highlight the last column (Example 5) with a pale yellow background to indicate the most generic case. 
  } 
  \label{fig:Histograms}
\end{figure*}

For each of the special cases, we then generate an ensemble of quantum trajectories with $R=5000$ realizations and analyze the trace squared-errors (TrSEs) compared to the true trajectories, which is $\sigma_{\text{A};k}^2(t_j)=2(1-|\bra{\psi_{\text{A};k}(t_j)}\psi_{\text{T};k}(t_j)\rangle|^2)$. Here, $\ket{\psi_{\text{A};k}(t_j)}$ denotes the quantum state generated by the map $\hat M_\text{A}$, and $\ket{\psi_{\text{T},k}(t_j)}$ is the corresponding true state. For better visualization, we further compute the time-average TrSEs over $N$ time steps for a trajectory, which is defined as
\begin{align}\label{TSE1}
    \sigma_{\text{A};k}^2&=\frac{1}{N}\sum_{j=1}^N\sigma_{\text{A};k}^2(t_j).
\end{align}
The distributions of $\sigma_{\text{A;}k}$ (square root of \erf{TSE1}) are presented as histograms in Fig.~\ref{fig:Histograms}. We then average $\sigma_{\text{A};k}$ over the ensemble $R$ realizations to obtain the MTrSE,
\begin{align}\label{TSE2}
    \bar\sigma_{\text{A}}&= \frac{1}{R}\sum_{k=1}^R \sigma_{{\text{A}};k}
\end{align}
where $\bar\sigma_{\text{A}}=\bar{\sigma}_{I_t}(\hat\psi_{\text{A};I_t})$ defined in Eq.~\eqref{MTrSE}. 

Finally, since the TrAE (trace distance) was used for the analytic results in the preceding section and in Ref.~\cite{WWC2024}, we also consider numerical values of $D_{\text{A}; k}(t_j)=\sqrt{1-|\bra{\psi_{\text{A};k}(t_j)}\psi_{\text{T};k}(t_j)\rangle|^2}$. From this we compute the time- and ensemble-\emph{mean trace-absolute error} (MTrAE), defined as 
\begin{align}
    \bar D_\text{A}&= \frac{1}{R}\sum_{k=1}^R \frac{1}{N}\sum_{j=1}^N D_{\text{A};k}(t_j).
\end{align}
This quantity enables a direct and consistent comparison between our numerical error analysis and the existing analytical results reported in Ref.~\cite{WWC2024}.

\subsection{Numerical results}\label{Num-results}

We show the numerical results for the square root of the time-average TrSEs shown as histograms for the five examples and all methods in Fig.~\ref{fig:Histograms}, along with the ensemble average of this quantity, and the corresponding average for the TrAE.

\sechead{Example 1} This example, where $\hat c=\sqrt{\gamma/2}\hat\sigma_z$, is presented in the first column. As noted above, this falls under Case~1, as $\hat c^2\propto \hat 1$, but also Case~4 as $[\hat c,\hat c^\dagger]=0$. As predicted by Eqs.~\eqref{case1-error}, the accuracy of map $\hat M_\text{I}$ becomes comparable to $\hat M_\text{R}$ and $\hat M_\text{W}$, with $\bar\sigma_\text{A}=O(10^{-3})$.  As predicted by Eqs.~\eqref{case4-error}, the accuracy of $\hat M_\text{\faFaucet}$ is much better, with $\bar\sigma\rb=O(10^{-5})$, a factor of $10^{-2}$ smaller. In terms of the MTrAE, it is smaller by a factor of about $10^{-2}$, exactly as expected from Eqs.~\eqref{case4-error} since we have $\gamma\Delta t = 10^{-2}$. 
We note that, although the pure state $\ket{\psi\rb}$ exhibits the square root of MTrSE with $\bar \sigma\rb=O[(\gamma\ddt)^{5/2}]$, 
due to the truncation of $\hat M\rb$ at $(\ddt)^2$, no such error would be expected for the exact $\rho\rb$ from Ref.~\cite{guilmin2025} because it is an instance of Case 4. In this case it appears that $\hat M_\Phi$ has comparable performance to that of $\hat M\rb$ with $O[(\ddt)^{5/2}]$, even though one would naively expect 
$\bar \sigma_\Phi=O[(\gamma\ddt)^{2}]$. Possible reasons for this behaviour are as follows. First, these scalings are accompanied by prefactors whose magnitudes are typically small, around $10^{-1}$. These prefactors were analytically derived in Ref.~\cite{WWC2024} (and we shall see some below). Second, it is possible that certain contributions at order $O[(\ddt)^2]$ vanish due to the normality of the operator $\hat c$. Verifying this speculation, however, would require an exact dynamical map evaluated to $O[(\ddt)^2]$.

Ref.~\cite{WWC2024} derived analytical expressions for the \emph{single-time-step} average distance between trajectories generated by the map $\hat{M}_{\mathrm{A}}$ and those generated by the \emph{exactly solvable} map, for Examples 1 and 2 and $\mathrm{A}\in \{\mathrm{I,R,W} \}$. These distances, denoted by $D_{\mathrm{A}}$, are reported in Table~II of Ref.~\cite{WWC2024} for Example~1. Specifically,
\begin{subequations}\label{D_Example1}
\begin{align}
    D_\text{I}&=0.2585(\gamma\ddt)^{3/2},\\
    D_\text{R}&=0.1551(\gamma\ddt)^{3/2},\\
    D_\text{W}&=0.0699(\gamma\ddt)^{3/2}.
\end{align} 
\end{subequations}
These analytical results provide a useful benchmark for validating our numerical simulations. Since the distances in Eqs.~\eqref{D_Example1} correspond to a single time step, whereas our numerical quantity $\bar{D}_{\mathrm{A}}$ is obtained from trajectories consisting of $N$ time steps, it is necessary to relate $D_{\mathrm{A}}$ and $\bar{D}_{\mathrm{A}}$. In Appendix~\ref{average-hist-derive}, we derive the relation
\begin{align}\label{av-tds}
    \bar D_\text{A}&=\frac{2}{3}\sqrt{N}D_\text{A}.
\end{align}
We note that $D_{\mathrm{A}}$ is averaged over the \emph{exact} probability distribution derived in Ref.~\cite{WWC2024}. Substituting Eqs.~\eqref{D_Example1} into Eq.~\eqref{av-tds}, with $\gamma\ddt=10^{-2}$ and $N=100$, yields values that are in good agreement with the numerical data (the values of $\bar D_{\mathrm{A}}$ shown in the first column). In particular, we obtain $\bar D_\text{A}=O[(\gamma\ddt)^{3/2}]=O(10^{-3})$ for $\text{A}\in\{\text{I,R,W}\}$. The small discrepancies between the analytical predictions and numerical results can be attributed to the averaging of $D_{\mathrm{A}}$ over the qubit initial state using the Haar measure in Ref.~\cite{WWC2024}.

\sechead{Example 2} In the second column of Fig.~\ref{fig:Histograms}, we present the qubit fluorescence measurement, $\hat c=\sqrt{\gamma}\hat\sigma_-$. 
As expected from \erf{case2-scaling}, we find that $\bar\sigma_\text{I}=\bar\sigma_\text{R}$ and $\bar\sigma_\text{W}=\bar\sigma\rb$, with these four maps producing the square root MTrSEs of order $O[(\gamma\ddt)^{3/2}]=O(10^{-3})$. The map $\hat M_\Phi$ now produces trajectories with error much smaller than those of $\hat M\rb$, because $\hat c$ is not normal. Note that, in this example, $D_\text{A}$ for A $\in \{\text{I, R, W}\}$ was also derived in Ref.~\cite{WWC2024}, reported in Table~II, with
\begin{align}
D_\text{I}=D_\text{R}&=0.1152(\gamma\ddt)^{3/2},\nonumber\\
D_\text{W}&=0.0576(\gamma\ddt)^{3/2}.\nonumber
\end{align}
We find that the corresponding numerical results $\bar D_\text{A}$ for A $\in\{\text{I, R, W}\}$ as displayed in the plots in the second column lead to an agreement in Eq.~\eqref{av-tds} by substituting the analytical results $D_\text{A}$ above.

\sechead{Example~3} The third column shows $\hat c=\sqrt{\gamma}\hat S_-$ for a spin-1 system. 
In this case, the It\^o map generates the largest error, with $\bar\sigma_\text{I}=O(\gamma\ddt)=O(10^{-2})$, while the error associated with $\hat M_\text{A}$ for $\text{A}\in\{\text{R,W,\faFaucet}\}$ now exhibit the same scaling as $\bar\sigma_\text{A}=O[(\ddt)^{3/2}]=O(10^{-3})$ with different prefactors. 
Notably, we see that $\bar\sigma_\text{W}\simeq\bar\sigma_\text{\faFaucet}$, both scaling as $O(10^{-3})$, in agreement with the prediction of Eq.~\eqref{ntdse}. 
This example is consistent with the behavior expected for Case~3.

\sechead{Example~4} The fourth column of Fig.~\ref{fig:Histograms} illustrates Case~4. where the operator $\hat c$ is normal, satisfying $[\hat c,\hat c^\dagger]=0$, but does not satisfy $\hat c^2 \propto \hat 1$ as in Case~1. This is achieved by taking $\hat c=\sqrt{\gamma/2}\hat S_z$ for a spin-1 system. 
As expected, the robinet method again produces the smallest MTrSE with its square root $\bar \sigma\rb = O[(\gamma\ddt)^{5/2}]=O(10^{-5})$. 
However, the $\hat M_\Phi$ still outperforms the other maps with scaling $\bar\sigma_\Phi=O[(\ddt)^2]=O(10^{-4})$, while $\bar\sigma_\text{W}, \bar\sigma_\text{R}=O[(\ddt)^{3/2}]=O(10^{-3})$ and $\bar\sigma_\text{I}=O(\ddt)=O(10^{-2})$.

\sechead{Example~5} Finally, for the fifth column of Fig.~\ref{fig:Histograms}, we take $\hat c=\sqrt{\gamma}\hat L_-$ for a spin-$3/2$ system. These results constitute the most important findings of this work and are highlighted with a pale-yellow background. This example corresponds to Case~5, the generic case. Here 
the map $\hat M_\Phi$ produces the smallest MTrSE, its square root scaling as $\bar \sigma_\Phi= O[(\ddt)^2]=O(10^{-4})$, while the other maps, $\hat M_\text{R}$, $\hat M_\text{W}$, and $\hat M_\text{\faFaucet}$, exhibit MTrSEs scaling as $O[(\gamma\ddt)^3]=O(10^{-6})$, with numerical prefactors decreasing from R to W to \faFaucet, as the number of terms in the map increases. Finally, the \ito\  map again has its square root MTrSE $\bar\sigma_\text{I}=O(\ddt)=O(10^{-2})$. It is worth emphasizing that, although we consider the case $\hat H=0$, the $\Phi$-map already outperforms all existing approaches in this generic setting. This is sufficient to demonstrate its superiority, and therefore it is unnecessary to consider additional scenarios involving a nonzero Hamiltonian $\hat H$.

\section{Conclusion}\label{conclusion}
We have analyzed the error in quantum trajectory simulations arising from finite experimental time resolution, which lead to quantum evolution conditioned only on time-binned measurement records $I_t$. Existing higher-order approaches—including the It\^o expansion, the maps of Rouchon--Ralph~\cite{Rouchon2015} and Wonglakhon \emph{et al.}~\cite{WWC2024}, and the robinet method~\cite{guilmin2025}—have deviations of order $O[(\ddt)^{3/2}]$ from the fully conditioned state. This is so even though the robinet map $\mathcal{K}\rb$ is optimal given the time-binned record $I_t$, because it is fundamentally limited by information lost in the time-binning process itself. 

This led us to derive what we call the $\Phi$-map, by returning again to the system--bath Hamiltonian formalism This map incorporates a single additional record variable, $\phi_t$, yet achieves trajectory estimates with an error scaling of order $(\ddt)^2$ relative to the fully conditioned state $\psiful$.  Using analytically solvable solutions, we quantified the typical TrAEs between $\psiful$ and the trajectories generated by existing methods for arbitrary measurement coupling operators $\hat c$, as well as measurement operators satisfying a variety of special conditions. 

Our analysis further reveals that the accuracy of $\mathcal{K}\rb$ is limited by non-QND effects where $[\hat H,\hat c]\ne 0$ or $[\hat c\dg,\hat c]\neq 0$. These non-commutativities give rise to evolution terms involving the additional record $\phi_t$ that cannot be neglected for higher-order accuracy. In contrast, for QND measurements where $\hat c$ is normal and satisfies $[\hat H,\hat c]=0$, we find that the variable $\phi_t$ vanishes from the evolution, and the robinet method becomes exact. 
We verified this analytic finding, and other special cases, with numerical simulations for a variety of finite-dimensional systems. 

Overall, the $\Phi$-map provides the highest accuracy presently available for general quantum trajectory estimation while requiring only a doubling of record variables, from $I_t$ alone to $(I_t, \phi_t)$. The record $\phi_t$ encodes higher-order temporal information within each measurement interval, thus capturing higher-order corrections to the quantum evolution. In contrast, the coarse-grained record $I_t$ discards this information. While $I_t$ is readily accessible in standard experiments, extracting $\phi_t$ would presumably require a dedicated electronic processing scheme to capture the linearly time-weighted raw data in experiments. Designing such schemes remains an open and promising direction for future research. Finally, we remind the reader that our $\Phi$-map is not the optimal map for the state conditioned on $(I_t, \phi_t)$. This means that it should be possible to derive a dynamical map, using those records, with error even smaller than $O[(\ddt)^2]$, 
by expanding to order $(\ddt)^2$ or beyond. However, it may involve many more terms, including cross-terms between $I_t$ and $\phi_t$ or even more complicated coarse-grained records beyond $\phi_t$.  We leave this extension for future work.

\section*{Acknowledgement}
This research was supported by the Australian Research Council Centre of Excellence Program [grant number CE170100012] and the Program Management Unit for Human Resources and Institutional Development, Research and Innovation (Thailand) [grant number B05F630108]. This research was supported in part by the International Centre for Theoretical Sciences (ICTS) for the program Quantum Trajectories (code: ICTS/qt2025/01).

\section*{Data availability}
The data that support the findings of this article are openly available~\cite{GitHubNTP}.
\appendix 

\section{Little and Big O notation}\label{O-Notations}
We provide the definitions of the little $o$ and big $O$ notations used throughout this paper to guide the reader.

\sechead{Little o notation} $f(x)=o(g(x))$ means: for every real positive constant $a$, there exists a real positive number $x_0$ such that for all $x>x_0$,
\begin{equation}
|f(x)| < a\,|g(x)|.
\end{equation}
Equivalently, $\lim_{x\to\infty} f(x)/g(x) = 0.$ That is, $f$ is asymptotically dominated by $g$.

\sechead{Big O notation} $f(x)=O(g(x))$ means: there exists a real positive constant $a$ and a real positive number $x_0$ such that for all $x>x_0$,
\begin{equation}
|f(x)| \le a\,|g(x)|.
\end{equation}
Equivalently, $\limsup_{x\to\infty} |f(x)|/|g(x)| < \infty.$ That is, $f$ is asymptotically bounded above by $g$ up to a constant factor.

In this work, we will also use these definitions in the limit $\ddt \to 0$, with the analogous interpretation. For instance, $f(\ddt)=o[(\ddt)^2]$ means that
\begin{equation}
|f(\ddt)| < a(\ddt)^2
\end{equation}
for sufficiently small $\ddt$, e.g., $f(\ddt)= a'(\ddt)^3$ for a constant $a'$. In contrast, $f(\ddt)=O[(\ddt)^2]$ means that
\begin{equation}
|f(\ddt)| \le a(\ddt)^2
\end{equation}
for sufficiently small $\ddt$, e.g., $f(\ddt)= a  (\ddt)^2 +a'(\ddt)^3$ for constants $a, a'$.

\section{Nearly exact map derivation}\label{NEM_derivation}
Following Eq.~\eqref{NEM_prod}, we construct the nearly exact map by keeping the deterministic terms to $O[(\delt)^2]$ and stochastic terms to $O[(\delt)^{3/2}]$. Hence, we obtain all possible terms as follows: 
\lipsum[0]
\begin{widetext}
\begin{multline}\label{NEM_expand}
    \hat M_{3/2}= \hat{1}+\hat{c}Y_t\ddt -\left[ \frac{1}{2}\hat{c}^\dagger\hat{c} -\frac{1}{2} \hat{c}^2\left(Y_t^2\ddt-1\right) +i\hat H\right]\ddt
    \\+ \lim_{m\rightarrow\infty} \Bigg\{ -\frac{1}{2}\hat c\hat c^\dagger\hat c \sum_{j=0}^{m-1}y_j\delt\left(j\delt+\frac{1}{3}\delt\right) -\frac{1}{2}\hat c^\dagger\hat c^2 \sum_{j=0}^{m-1}y_j\delt\left[(m-j)\delt+\frac{2}{3}\delt\right] -i\hat c\hat H \sum_{j=0}^{m-1}y_j\delt\left(j\delt+\frac{1}{2}\delt\right) \\
    -i\hat H\hat c \sum_{j=0}^{m-1}y_j\delt\left[(m-j)\delt+\frac{1}{2}\delt\right]
    +\hat c^3 \bigg[  \sum_{j=0}^{m-1}y_j\delt\sum_{k=0}^{j-1}\frac{1}{2}(y_k^2\delt-1)\delt +\sum_{j=0}^{m-1} \frac{1}{2}(y_j^2\delt-1)\delt\sum_{k=0}^{j-1}y_k\delt\\
    +  \sum_{j=0}^{m-1}y_j\delt \sum_{k=0}^{j-1}y_k\delt \sum_{\ell=0}^{k-1}y_\ell\delt
    + \sum_{j=0}^{m-1} \frac{1}{6}(y_j^3\delt-3y_j)\delt^2 \bigg]
    \Bigg\}.
\end{multline}
\end{widetext}
\lipsum[0]

For the first line in RHS of Eq.~\eqref{NEM_expand}, the first three terms are trivial and have been shown in Ref.~\cite{WWC2024}. 

For the second line and the first term of the third line of Eq.~\eqref{NEM_expand}, these terms are the exchange of $y_s \hat c\delt \times \hat R$, $\hat 1 \times Ay_s\hat c\hat R\delt^2$ and $\hat 1 \times Ay_s\hat R \hat c^\dagger\delt^2$ with $\hat R= -i\hat H\delt-\frac{1}{2}\hat c^\dagger\hat c\delt$ for a relevant coefficient $A$. Here, we can express the relevant four forms of the infinite sums as follows:
\begin{subequations}\label{correct-prefactors}
\begin{align}
     &\lim_{m\rightarrow\infty}\sum_{j=0}^{m-1}y_j\delt\left(j\delt+A\delt\right)\nonumber\\
     &=\int_{t}^{t+\ddt}\dd sy_s(s-t)\nonumber \\
     &=\int_{t}^{t+\ddt}\dd sy_s\left[\frac{\ddt}{2} + s-(t+\frac{\ddt}{2})\right]\nonumber\\
     &= \frac{1}{2}\ddt^2Y_t + Z_t,\\
     &\lim_{m\rightarrow\infty}\sum_{j=0}^{m-1}y_j\delt\left[(m-j)\delt+A\delt\right]\nonumber\\
     &=\int_{t}^{t+\ddt}\dd sy_s[\ddt-(s-t)]\nonumber\\
     &=\int_{t}^{t+\ddt}\dd sy_s\left[\frac{\ddt}{2} - s+(t+\frac{\ddt}{2})\right]\nonumber\\
     &= \frac{1}{2}\ddt^2Y_t - Z_t,
\end{align}
\end{subequations}
which are in terms of $Y_t$ and $Z_t$ records as defined in the main text. Thus, we have $\frac{1}{2}Y_t\ddt^2\{\hat c, \hat R\}+Z_t[\hat c,\hat R]=-\frac{1}{2}Y_t\ddt^2\{\hat c, i\hat H+\frac{1}{2}\hat c^\dagger\hat c\}- Z_t[\hat c,i\hat H+\frac{1}{2}\hat c^\dagger\hat c]$. We note that there are errors in Appendix~C of Ref.~\cite{WWC2024}. The prefactors for $\hat c\hat c^\dagger\hat c$ and $\hat c^\dagger\hat c^2$ in Eqs.~(C7) and~(C8) should follow Eqs.~\eqref{correct-prefactors} above. However, this does not affect the results in Ref.~\cite{WWC2024} as they only focus on the ensemble average evolution to $(\ddt)^2$, while the contribution from $Z_t$ appears in $(\ddt)^3$.

The rest of the terms in Eq.~\eqref{NEM_expand} are proportional to $\hat c^3$, which are a result of four possible exchanges: (a) $y_{s}\hat c\delt \times  \tfrac{1}{2}(y_{s'}^2\delt-1)\hat c^2\delt$, (b)  $ \tfrac{1}{2}(y_s^2\delt-1)\hat c^2\delt \times  y_{s'}\hat c\delt$, (c)  $y_{s}\hat c\delt \times y_{s'}\hat c\delt \times y_{s''}\hat c\delt$, and (d) $\hat 1 \times \tfrac{1}{6}(y_{s}^3\delt-3y_{s})\delt^2\hat c^3$.

For the exchange (a), we can write the relevant form:
    \begin{align}
       \sum_{j=0}^{m-1}y_j\delt\sum_{k=0}^{j-1}\tfrac{1}{2}(y_k^2\delt-1)\delt&=\tfrac{1}{2}\sum_{j>k=0}^{m-1} y_j\delt(y_k^2\delt-1)\delt.\label{config_a}
    \end{align}
Likewise the exchange (b), we write:
    \begin{align}
       \sum_{j=0}^{m-1} \tfrac{1}{2}(y_j^2\delt-1)\delt\sum_{k=0}^{j-1}y_k\delt&=\tfrac{1}{2}\sum_{j>k=0}^{m-1} (y_j^2\delt-1)\delt y_k\delt.\label{config_b}
    \end{align}
Here, we combine Eq.~\eqref{config_a} and Eq.~\eqref{config_b}, which is simplified to:
    \begin{align}
        &\tfrac{1}{2}\sum_{j>k=0}^{m-1}y_j\delt(y_k^2\delt-1)\delt +  \tfrac{1}{2}\sum_{j>k=0}^{m-1}(y_j^2\delt-1)\delt y_k\delt\nonumber\\
        &=\sum_{j=0}^{m-1}y_j\delt\sum_{k=0}^{m-1}\tfrac{1}{2}(y_k^2\delt-1)\delt-\sum_{j=0}^{m-1}\tfrac{1}{2}y_j(y_j^2\delt-1)\delt^2\nonumber\\
        &=\frac{1}{2}\ddt Y_t\sum_{j=0}^{m-1}(y_j^2\delt-1)\delt-\tfrac{1}{2}\sum_{j=0}^{m-1}y_j(y_j^2\delt-1)\delt^2.\label{com_ab}
    \end{align}
For the exchange (c), we express:
    \begin{align}\label{threey}
       \sum_{j=0}^{m-1}y_j\delt \sum_{k=0}^{j-1}y_k\delt \sum_{\ell=0}^{k-1}y_\ell\delt &= \sum_{j>k>\ell =0}^{m-1} y_jy_ky_\ell\delt^3.
    \end{align}
By using the fact that
\begin{multline}
        \ddt^3Y_t^3=\sum_{j,k,\ell=0}^{m-1}y_jy_ky_\ell\delt^3\\
        =\sum_{j=0}^{m-1}y_j^3\delt^3+3\sum_{j\ne k=0}^{m-1}y_j^2y_k\delt^3+6\sum_{j>k>\ell=0}^{m-1}y_jy_ky_\ell\delt^3,\nonumber
    \end{multline}
we can be reformulated Eq.~\eqref{threey} as
\lipsum[0]
\begin{widetext}
\begin{align}
     \sum_{j>k>\ell=0}^{m-1}y_jy_ky_\ell\delt^3 &=\frac{1}{6}\left[\ddt^3Y_t^3-\sum_{j=0}^{m-1}y_j^3\delt^3-3\sum_{j\ne k=0}^{m-1}y_j^2y_k\delt^3\right]=\frac{1}{6}\ddt^3Y_t^3-\frac{1}{6}\sum_{j=0}^{m-1}y_j^3\delt^3
    -\frac{1}{2}\left(\sum_{j,k=0}^{m-1}y_j^2y_k\delt^3-\sum_{j=0}^{m-1}y_j^3\delt^3\right)\nonumber \\
        &=\frac{1}{6}\ddt^3Y_t^3-\frac{1}{6}\sum_{j=0}^{m-1}y_j^3\delt^3
        -\frac{1}{2}\left(\ddt Y_t\sum_{j=0}^{m-1}y_j^2\delt^2-\sum_{j=0}^{m-1}y_j^3\delt^3\right)\label{sim_c}.
\end{align}
Lastly, for the exchange (d), we have
    \begin{align}
       \sum_{j=0}^{m-1} \frac{1}{6}(y_j^3\delt-3y_j)\delt^2.\label{high_d}
    \end{align}
Thus, combining Eqs.~\eqref{com_ab}, ~\eqref{sim_c} and~\eqref{high_d}, we find the terms that is proportional to $\hat c^3$ as
 \begin{multline}
     \frac{1}{2}\ddt Y_t\sum_{j=0}^{m-1}(y_j^2\delt-1)\delt-\frac{1}{2}\sum_{j=0}^{m-1}y_j(y_j^2\delt-1)\delt^2+\frac{1}{6}\ddt^3Y_t^3-\frac{1}{6}\sum_{j=0}^{m-1}y_j^3\delt^3
     \\
     -\frac{1}{2}\left(\ddt Y_t\sum_{j=0}^{m-1}y_j^2\delt^2-\sum_{j=0}^{m-1}y_j^3\delt^3\right)+\sum_{j=0}^{m-1} \frac{1}{6}(y_j^3\delt-3y_j)\delt^2
     = \frac{1}{6}\ddt^3Y_t^3-\frac{1}{2}\ddt^2 Y_t.
 \end{multline}
 \end{widetext}
\lipsum[0]
Finally, the coefficient of $\hat c^3$ reads: $\frac{1}{6}(\ddt Y_t^3-3 Y_t)\ddt^2\hat c^3$ and obtain the closed form as
\begin{multline}\label{NEM_YZ}
     \hat M_{3/2}(Y_t, Z_t)=\hat{1}-\tfrac{1}{2}\hat{c}^\dagger\hat{c}\ddt+\hat{c}\ddt Y_t+\tfrac{1}{2}\hat{c}^2[Y_t^2\ddt^2-\ddt] \\ +\tfrac{1}{6}\ddt^2\hat c^3(Y_t^3\ddt-3Y_t) -\tfrac{1}{2}\left\{\hat{c}, i\hat H+\tfrac{1}{2}\hat{c}^\dagger\hat{c}\right\}\ddt^2Y_t\\ -Z_t\left[\hat{c},i\hat H+ \tfrac{1}{2}\hat{c}^\dagger\hat{c}\right] +\tfrac{1}{2}\left[\left(\tfrac{1}{2}\hat{c}^\dagger\hat{c}\right)^2-\hat H^2\right]\ddt^2,
\end{multline}
where the variable $Z_t$ was first introduced in Ref.~\cite{WWC2024}, defined as
\begin{align}
     Z_t&\equiv \int_t^{t+\ddt}\dd s y_s\left[s-\left(t+\frac{\ddt}{2}\right)\right].
\end{align}
The variable $Z_t$ also has a Gaussian ostensible probability distribution, given by
\begin{align}\label{ostZ}
\wp_\text{ost}(Z_t)&=\sqrt{\frac{6}{\pi \ddt^3}}\exp(-6Z_t^2/\ddt^3).
\end{align}
It has been shown that $Y_t$ and $Z_t$ are statistically independent, i.e.,
 $\text{E}[Y_t Z_t]_{y_s} = 0$, where we have used the ensemble average notation: $\text{E}[\bullet]_{y_t}\equiv \int_{-\infty}^\infty\dd y_t \wp_\text{ost}(y_t)\bullet$. Given the independence of $Y_t$ and $Z_t$, the Kraus form of the nearly exact map reads $\hat K_{3/2}(Y_t,Z_t)=\sqrt{\wp_\text{ost}(Y_t)\wp_\text{ost}(Z_t)}\hat M_{3/2}(Y_t,Z_t)$.

\section{ Mixed state formalism for quantum trajectory simulations}\label{MixedStateQTJ}
When the measurement efficiency is not unity, i.e., $\eta<1$, the quantum state evolution loses its purity and becomes mixed. Hence, we instead use $\sqrt{\eta}\hat c$ in the measurement operators and employ the density matrix formalism to describe the quantum evolution, giving
\begin{align}
    \rho(t+\ddt)
    =
    \frac{\mathcal{K}_{\rm A}^\eta\left[\rho(t)\right]}
    {\mathrm{Tr}\!\left\{
    \mathcal{K}_{\rm A}^\eta\left[\rho(t)\right]
    \right\}},
\end{align}
where we have introduced the superoperator $\mathcal{K}_{\rm A}^\eta$ to accommodate the additional dephasing arising from the \emph{unmeasured} channel with $\text{A}\in \{\text{I, R, W, \faFaucet,} \Phi\}$ as defined in the main text. Here,
\begin{align}
    \mathcal{K}_{\text{A}}^\eta
    =
    \mathcal{K}_\text{A}(\sqrt{\eta}\hat c)
    +\ddt\mathcal{D}[\sqrt{1-\eta}\hat c].
\end{align}

We remark that, for the nearly exact map, the superoperator $\mathcal{K}_\Phi^\eta$ is accurate to $O[(\ddt)^{3/2}]$, determined by $\mathcal{K}_\Phi$. This is because the next-order term in the expansion of the unmeasured evolution is $\frac{1}{2}(\ddt)^2\mathcal{D}^2$, and the overall accuracy is therefore dictated by $\mathcal{K}_\Phi$.

\section{Derivation for the overall average TrAE}\label{average-hist-derive}

The goal of this derivation is to show the relationship between the accumulating $N$ time steps TrAE with a single time step TrAE from the known results in Ref~\cite{WWC2024}. 

Let us first consider a Gaussian random variable $Q$, featuring the probability distribution $\wp_Q$ with $\mu_Q$ mean and $\sigma_Q^2$ variance. Here, we define the absolute distance between the random variable $q$ and its mean for one time step as $D_q\equiv |q-\mu_Q|$. The distance average is given by 
\begin{align}
\text{E}\left[D_q\right]_{Q}=&\sqrt{\frac{2}{\pi}}\sigma,
\end{align}
and we can re-express the variance via $\sigma^2=\frac{\pi}{2} \text{E}^2[D_q]_{Q}$. Assuming the variance accumulates linearly over time, we can express the variance after $k$-th time step as 
\begin{align}
k\sigma^2&=\frac{k\pi}{2}\text{E}^2[D_q]_{Q}.
\end{align} 
Hence, we obtain the relation: $\sqrt{(\frac{2}{\pi}k)}\sigma=\sqrt{k}\text{E}[D_q]_{Q}$, for the absolute distance at $k$-th time-steps. We can compute the average overall distance after $K$ time steps, given by
\begin{align}
    \bar D&=\frac{1}{K}\sum_{k=0}^K \sqrt{k}\text{E}[D_q]_{Q}.
\end{align}
For a large number $K$, we can replace the summation to an integral as
\begin{align}\label{trace-hist-av}
\bar D&= \frac{\text{E}[D_q]_{Q}}{K}\int_{0}^K \dd k\sqrt{k}= \frac{2}{3}\sqrt{K}\text{E}[D_q]_{Q}.
\end{align}

As we already have the analytical results for $\text{E}[D_q]_{Q} \rightarrow D_\text{A}$ from Ref.~\cite{WWC2024} (for the qubit cases) and in our simulation we use $K\rightarrow N$, we have the overall average TrAE after $N$ time steps are given by
\begin{align}
    \bar D_\text{A}&=\frac{2}{3}\sqrt{N}D_\text{A}.
\end{align}

\bibliographystyle{apsrev4-1}
%

\end{document}